\documentclass[conference]{IEEEtran}
\usepackage{amsmath,amsfonts}
\usepackage{algorithmic}
\usepackage{algorithm}
\usepackage{array}
\usepackage[caption=false,font=normalsize,labelfont=sf,textfont=sf]{subfig}
\usepackage{textcomp}
\usepackage{stfloats}
\usepackage{url}
\usepackage{verbatim}
\usepackage{graphicx}
\usepackage{longtable}
\usepackage{cite}
\usepackage{hyperref}
\usepackage{multirow}
\usepackage{booktabs}
\usepackage[most]{tcolorbox}
\usepackage{array} 
\usepackage{enumitem}
\usepackage{ragged2e} 
\usepackage{booktabs} 
\usepackage{caption} 
\usepackage{hyperref} 
\usepackage{lipsum} 
\usepackage{caption}
\usepackage{subcaption}
\usetikzlibrary{mindmap, positioning}
\usepackage[margin=1in]{geometry}

\usepackage{tikz}
\usetikzlibrary{mindmap, shadows, positioning, backgrounds}
\usetikzlibrary{positioning, shapes, arrows.meta} 
\usetikzlibrary{mindmap, shadows, positioning, backgrounds}
\usetikzlibrary{positioning, shapes, arrows.meta}

\definecolor{scientificblue}{RGB}{0,82,147}
\definecolor{scientificgreen}{RGB}{0,132,61}
\definecolor{scientificpurple}{RGB}{94,54,137}
\definecolor{scientificorange}{RGB}{227,114,34}
\definecolor{scientificred}{RGB}{186,12,47}

\definecolor{lightblue}{RGB}{103,153,200}
\definecolor{lightgreen}{RGB}{132,199,143}

\usepackage{booktabs}
\usepackage{caption}
\begin{document}

\title{Vibe Coding vs. Agentic Coding: Fundamentals and Practical Implications of Agentic AI}

\author{
    Ranjan Sapkota\IEEEauthorrefmark{1}\IEEEauthorrefmark{3},
    Konstantinos I. Roumeliotis\IEEEauthorrefmark{2},
    Manoj Karkee\IEEEauthorrefmark{1}\IEEEauthorrefmark{3}
    \\
    \IEEEauthorblockA{\IEEEauthorrefmark{1}Cornell University, Department of Biological and  Environmental Engineering, USA}
    
    \IEEEauthorblockA{\IEEEauthorrefmark{2}University of the Peloponnese, Department of Informatics and Telecommunications, Tripoli, Greece}
    \\
    \IEEEauthorblockA{\IEEEauthorrefmark{3}Corresponding authors: \texttt{rs2672@cornell.edu}, \texttt{mk2684@cornell.edu}}
}

\maketitle
\begin{abstract}
This review presents a comprehensive analysis of two emerging paradigms in AI-assisted software development: \textit{vibe coding} and \textit{agentic coding}. While both leverage large language models (LLMs), they differ fundamentally in autonomy, architectural design, and the role of the developer. Vibe coding emphasizes intuitive, human-in-the-loop interaction through prompt-based, conversational workflows that support ideation, experimentation, and creative exploration. In contrast, agentic coding enables autonomous software development through goal-driven agents capable of planning, executing, testing, and iterating tasks with minimal human intervention. We propose a detailed taxonomy spanning conceptual foundations, execution models, feedback loops, safety mechanisms, debugging strategies, and real-world tool ecosystems. Through comparative workflow analysis and 20 detailed use cases, we illustrate how vibe systems thrive in early-stage prototyping and education, while agentic systems excel in enterprise-grade automation, codebase refactoring, and CI/CD integration. We further examine emerging trends in hybrid architectures, where natural language interfaces are coupled with autonomous execution pipelines. Finally, we articulate a future roadmap for agentic AI, outlining the infrastructure needed for trustworthy, explainable, and collaborative systems. Our findings suggest that successful AI software engineering will rely not on choosing one paradigm, but on harmonizing their strengths within a unified, human-centered development lifecycle.
\end{abstract}



\begin{figure}
    \centering
    \includegraphics[width=\linewidth]{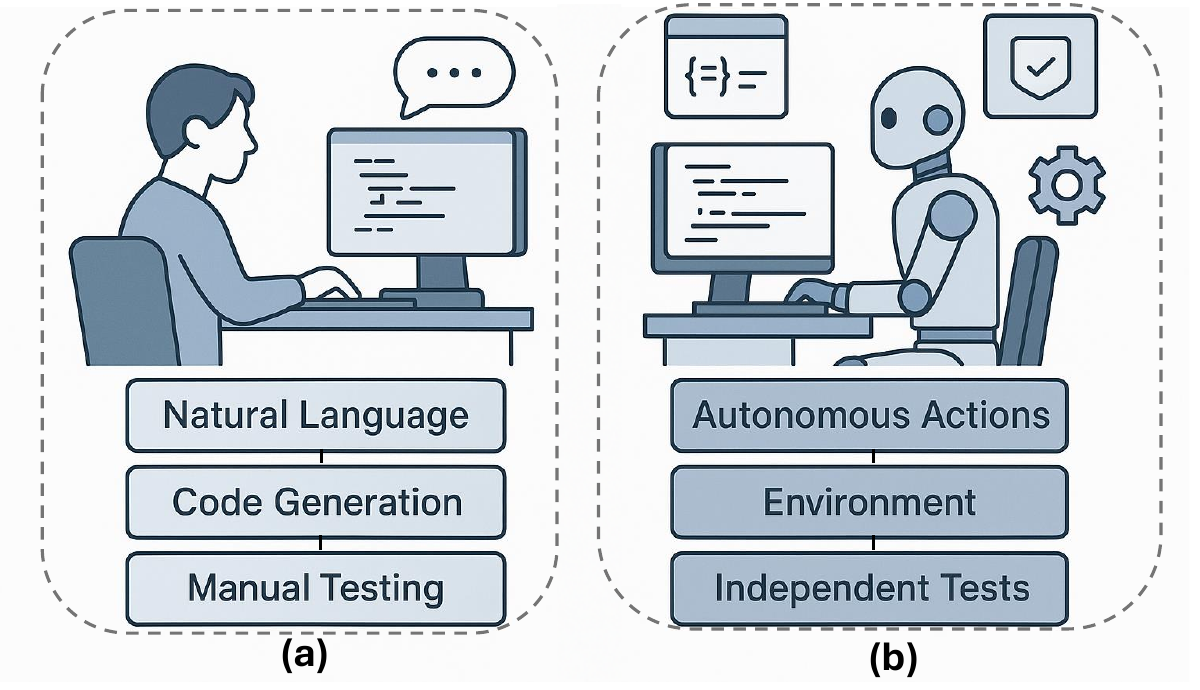}
    \caption{\textbf{Bird-eye-figure} showing a comparative illustration of \textbf{(a)} \textit{Vibe Coding}, where a human developer uses Natural Language to guide Code Generation followed by Manual Testing, and \textbf{(b)} \textit{Agentic Coding}, where an AI agent performs Autonomous Actions within an execution Environment, enabling Independent Tests. This highlights key differences in input modality, execution autonomy, and testing workflow between the two paradigms.}
    \label{fig:birdeye}
\end{figure}
\section{Introduction}
“Vibe coding” refers to a novel, emergent mode of software development in which the human programmer (\ref{fig:birdeye}a) operates less as a direct implementer of code and more as a high-level coordinator who collaborates with LLMs through iterative prompting and strategic direction \cite{chow2025technology}. Coined by Andrej Karpathy 
 \href{https://en.wikipedia.org/wiki/Andrej_Karpathy}{Andrej Karpathy Wiki Page} ,\href{https://www.technologyreview.com/2025/04/16/1115135/what-is-vibe-coding-exactly/}{1}, \href{https://en.wikipedia.org/wiki/Vibe_coding}{2}, the term captures a shift in both mindset and methodology, where developers communicate desired outcomes the “vibe” via natural language instructions, conceptual overviews, and progressive refinements, rather than by specifying logic in syntactic detail. Unlike traditional paradigms, which emphasize mastery over syntax \cite{weise1993programmable, mccane2017mastery, luxton2018developing} and low-level operations\cite{condit2007dependent, protzenko2017verified, balliu2014automating}, vibe coding reorients the focus toward intent specification, architectural vision, and interactive debugging \cite{chow2025technology}. At its core, vibe coding integrates principles from prompt engineering, agile design, and human-AI co-creation, while abstracting much of the linguistic burden onto the LLM \cite{gadde2025democratizing}. 
 
This interactional loop of guidance, AI response, human evaluation, and corrective feedback yields a dynamic coding process that is simultaneously expressive, generative, and improvisational. It invites the question: \textbf{can the act of software engineering become more intuitive, collaborative, and aligned with human reasoning, rather than merely a transcription of formal logic into text?} Vibe coding attempts to answer in the affirmative proposing a new semiotic contract between the human mind and generative machines.

The rise of vibe coding parallels the rapid advancement of foundation models and the growing availability of LLM-based development platforms like ChatGPT \cite{achiam2023gpt}, Replit \href{https://replit.com/ai}{ Replit AI}, and Cursor \href{https://www.cursor.com/en}{CURSOR}. Traditional software engineering emphasized rigid syntax\cite{fiadeiro2005categories, givon2011iconicity, maiorana2015quizly}, algorithmic structure \cite{tamburri2013organizational, dourish2016algorithms}, and deterministic logic\cite{gordon2014probabilistic, jenson2016deterministic}. In contrast, LLMs now allow developers to produce coherent, context-aware code using natural language transforming code creation into a dialogue with the machine \cite{liskavets2025prompt, li2024enhancing}. Karpathy’s notion of “embracing exponentials” reflects this paradigm shift, where scale, abstraction, and expressiveness redefine programming practice \href{https://karpathy.ai/}{Andrej Karpathy}, \href{https://itpma.notion.site/GPT-from-scratch-A-Karpathy-cd8c70028b1148cda2eb961141b50bcf}{ Notion}.

While vibe coding represents a leap in developer productivity and human-AI interaction, agentic coding signifies a more advanced and autonomous evolution of AI-assisted programming. At its core, agentic coding (Figure \ref{fig:birdeye}b) is grounded in the deployment of agentic AI systems software agents capable of independently interpreting high-level goals, decomposing tasks into subtasks, planning execution strategies, and adapting behavior based on real-time feedback and outcomes\cite{sapkota2025ai, pohler2025keeping}. Unlike the prompt-response dynamic of vibe coding, agentic coding minimizes the need for continuous human oversight. It introduces a paradigm wherein AI agents can initiate action, access tools and APIs, retrieve and process external data, and iteratively refine outputs through cycles of self-evaluation. These agents exhibit hallmark capabilities of agency, including intentionality, forethought, and adaptivity aligning with definitions proposed in cognitive systems and artificial general intelligence research. This level of autonomy enables agentic coding systems to tackle complex, multi-step workflows, making them suitable for process automation, business operations, and dynamic, data-driven environments. Architecturally, agentic coding employs reinforcement learning and modular planning \cite{borghoff2025human}, often integrating specialized subagents that collaborate to complete broader missions \cite{bousetouane2025agentic}. In contrast to vibe coding’s focus on expressiveness and flow, agentic coding is outcome-oriented, resilient, and self-directed. If vibe coding equips developers with a high-speed copilot, agentic coding gives them an intelligent collaborator capable of independently steering the aircraft. This distinction underscores not merely a difference in tools, but a fundamental rethinking of how software is authored, executed, and evolved.

Understanding the distinction between vibe coding and agentic coding is crucial for navigating the future landscape of AI-assisted software development. These paradigms differ not just in their technical mechanisms, but in the underlying assumptions about the role of the human developer, the locus of agency, and the nature of control in the coding process. Vibe coding maintains a human-centric model, where the developer is an ever-present conductor guiding the AI’s output a model well-suited to creative ideation, exploratory development, and rapid iteration. Agentic coding, on the other hand, introduces AI agents as semi-autonomous collaborators, shifting the human role to that of a supervisor who defines goals and evaluates outcomes. This shift has profound implications: it reconfigures workflows, challenges traditional notions of authorship and accountability, and necessitates new interfaces for monitoring, debugging, and aligning agent behavior with human intent. Moreover, the increasing prevalence of agentic systems raises critical questions around safety, reliability, and trustworthiness in software generated or managed with minimal human oversight. 

As organizations, developers, and researchers embrace these technologies, a dynamic understanding of the boundaries, affordances, and use cases of each paradigm becomes essential. By developing a formal taxonomy of vibe coding and agentic coding as this paper aims to do we not only chart the evolution of AI in programming but also lay the groundwork for better tooling, clearer expectations, and more robust human-AI collaboration in both current and future systems.

\section{Conceptual Foundations}
\subsection{Vibe Coding: Intuition-Driven, Developer-Centric Code Generation}
Vibe Coding, a term popularized by Andrej Karpathy \href{https://karpathy.ai/}{Andrej Karpathy}, describes a software development methodology centered on the developer's intuitive expression of intent to an LLM, which then acts as a highly responsive co-pilot in generating code. The "vibe" refers to the desired outcome, functionality, or even the aesthetic feel of a software component, which the developer communicates, often through natural language, for the LLM to translate into executable code. This paradigm fundamentally alters the traditional coding process by abstracting significant portions of syntactical detail and boilerplate, allowing developers to focus on higher-level design and rapid iteration.

\subsubsection{The Semiotics of Vibe Coding}
At its core, Vibe Coding introduces a new semiotic layer in programming \cite{yuan2025vibe, yin2025vibes}. Traditionally, developers use programming languages as a direct, formal means of instructing a computer \cite{de1991learning, maurer2006theory, smetana2012computer}. In Vibe Coding, natural language, high-level descriptions, and even examples or visual mock-ups serve as primary inputs to an intermediary intelligent system (the LLM) \cite{nam2024using}. The LLM, in turn, interprets these multi-modal "signs" and synthesizes corresponding formal code \cite{liao2024gpt, coignion2024performance}. This process is not a one-way translation but an iterative dialogue. The developer provides an initial prompt; the LLM generates a code artifact; the developer reviews, critiques, and refines the prompt or directly edits the code, continuing the cycle until the desired "vibe" is achieved \cite{rasheed2024ai, wang2023review}. This iterative refinement loop is a defining characteristic, reflecting a co-constructive model of meaning-making between human and machine \cite{fichtel2025investigating}.

\subsubsection{Fundamental Skills and Cognitive Shifts}
\begin{figure}[ht!]
    \centering
    \includegraphics[width=\linewidth]{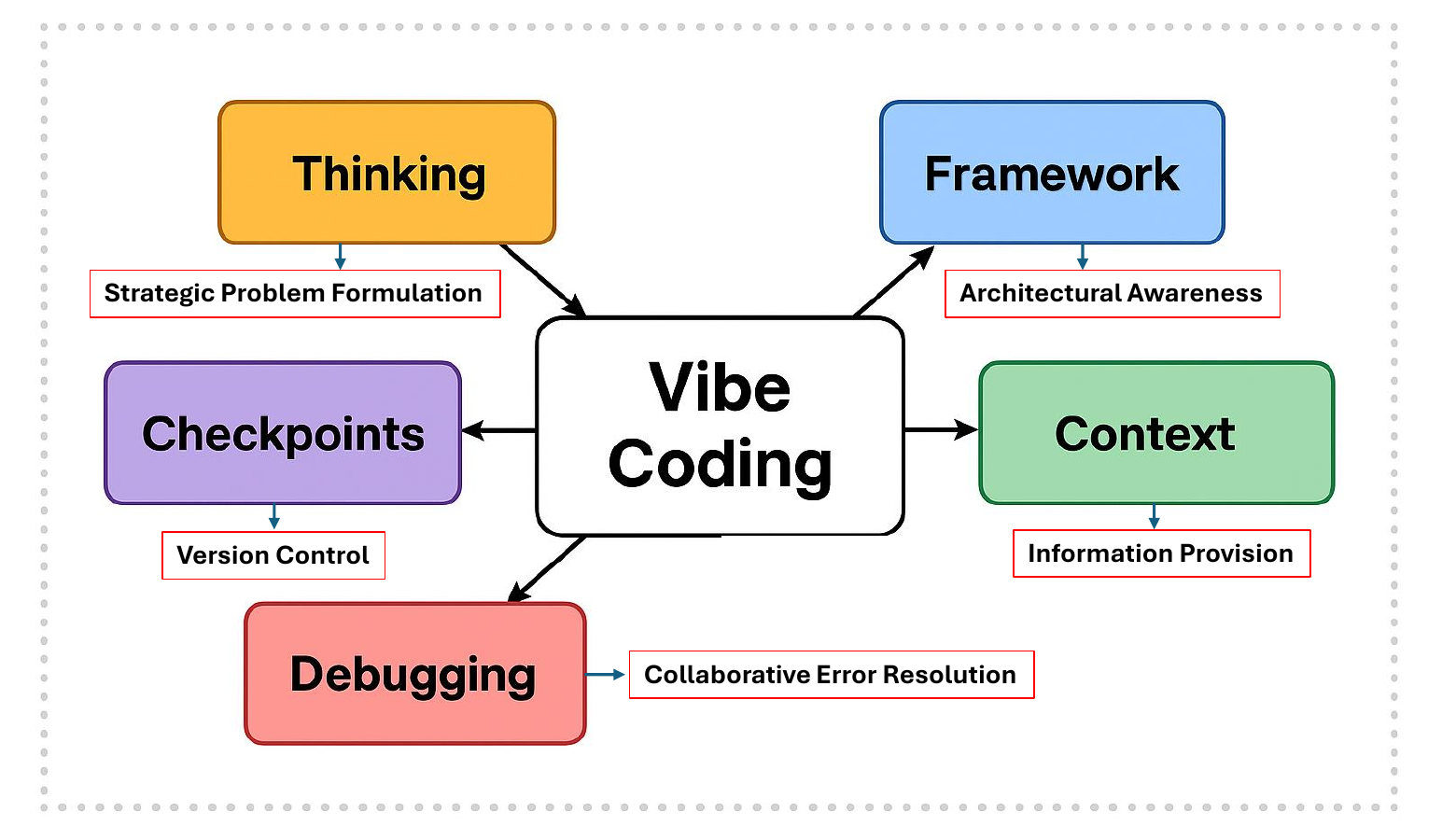}
    \caption{Fundamental Skills and Cognitive Shifts in Vibe Coding. This diagram illustrates the five core competencies Thinking, Framework, Checkpoints, Debugging, and Context that enable effective collaboration with LLMs. Together, they represent a cognitive shift from syntax-heavy implementation to high-level conceptual guidance and iterative co-creation with AI agents.}
    \label{fig:vibecoding}
\end{figure}

Philosophically, Vibe Coding empowers the developer by augmenting their cognitive capabilities. It offloads the burden of low-level implementation details, allowing for a greater focus on creative problem-solving, user experience design, and system architecture. The "vibe" is thus not merely an aesthetic preference but a holistic representation of the developer's intent, encompassing functionality, usability, and design. Effective Vibe Coding necessitates a shift in developer skills, emphasizing conceptual articulation and strategic interaction over rote memorization of syntax \cite{mesko2023prompt, white2023prompt}. The five fundamental skills are as illustrated in Figure \ref{fig:vibecoding} and each skills are explained below:
\begin{itemize}
    \item \textbf{Thinking (Strategic Problem Formulation):} This involves a multi-layered approach to defining the problem for the LLM. It begins with \textit{Logical Thinking} (the core what) \cite{chang2023prompting, alfageeh2025prompts}, progresses to \textit{Analytical Thinking} (how users interact, high-level components) \cite{xiao2024enhancing, gao2024taxonomy, ma2024beyond}, then to \textit{Computational Thinking} \cite{yan2025llm} (structuring the problem into modules, rules, and data flows understandable by the AI) \cite{xenakis2025llm, yang2025code}, and finally to \textit{Procedural Thinking} (considering optimal execution, best practices, and detailed features) \cite{subramonyam2024bridging, wu2024thinking}. A well-crafted Product Requirements Document (PRD) often emerges from this rigorous thinking process, serving as a detailed contextual blueprint for the LLM.
    
    \item \textbf{Framework (Architectural Awareness):} While the LLM handles much of the implementation, the developer must possess an awareness of relevant software frameworks (e.g., React, Node.js, Django), libraries, and architectural patterns \cite{xu2025powerpoint}. This knowledge allows the developer to guide the LLM towards using appropriate, robust, and industry-standard technologies, thereby constraining the solution space and improving code quality and maintainability \cite{yigit2025generative}. The developer can also learn about new frameworks by querying the LLM for recommendations based on project requirements.
    
    \item \textbf{Checkpoints (Version Control):} Given the generative and sometimes unpredictable nature of LLM outputs, robust version control (e.g., Git) is paramount. Frequent commits create "save points," enabling developers to revert to stable states if AI-generated code introduces errors or undesirable changes. Branching allows for safe experimentation with different AI-generated features without impacting the main codebase \cite{wei2023jailbroken, mou2024sg}. This ensures a safety net for the rapid, iterative cycles inherent in Vibe Coding.
    
    \item \textbf{Debugging (Collaborative Error Resolution):} Errors are inevitable. In Vibe Coding, debugging becomes a collaborative process \cite{p2025bugspotter, jiang2024ledex}. The developer identifies an issue (runtime error, logical flaw, UI discrepancy) and then provides the LLM with rich context error messages, relevant code snippets, descriptions of expected vs. actual behavior, and sometimes screenshots \cite{tian2024debugbench, lee2024unified}. The LLM can then assist in diagnosing the problem and suggesting or implementing fixes. Human oversight is critical to guide this process and validate the AI's solutions.
    
    \item \textbf{Context (Information Provision):} The efficacy of Vibe Coding is directly proportional to the quality and comprehensiveness of the context provided to the LLM \cite{nam2024using, an2024make}. This includes not only the initial PRD and prompts but also visual mock-ups, examples of desired output, existing codebase snippets, API documentation for integrations, and explicit statements about preferred libraries, coding styles, or security constraints \cite{liu2024we, kapania2025m}. Rich context minimizes ambiguity and helps the LLM generate more accurate and relevant code.
\end{itemize}

\subsubsection{Interaction Model and Workflow Integration}
The interaction model in Vibe Coding is predominantly a tight, iterative prompt-response loop. The developer initiates with a high-level request, the LLM generates code, the developer reviews and refines either by editing the code directly or by providing a new, more specific prompt \cite{hou2024large, lyu2024automatic}. This cycle repeats, often rapidly, enabling quick prototyping and exploration of different solution paths.

Vibe Coding tools, such as AI-enhanced IDEs (e.g., Cursor \href{https://www.cursor.com/}{CURSOR}, Windsurf \href{https://windsurf.com/}{Windsurf}) or cloud-based platforms (e.g., Replit), integrate into the developer's workflow by providing an interface for this interaction. However, the execution and final validation of the generated code typically occur within a standard development environment, often managed by the developer. This separation between generation and execution necessitates careful testing and integration, as the LLM does not inherently possess a runtime understanding of the code it produces in most Vibe Coding scenarios. This model thrives in creative and exploratory development phases but requires disciplined application of checkpointing and refactoring to manage potential technical debt accrued from rapid, less scrutinized code generation.

\subsection{Agentic Coding: Towards Autonomous Software Development Systems}

\begin{figure}
    \centering
    \includegraphics[width=\linewidth]{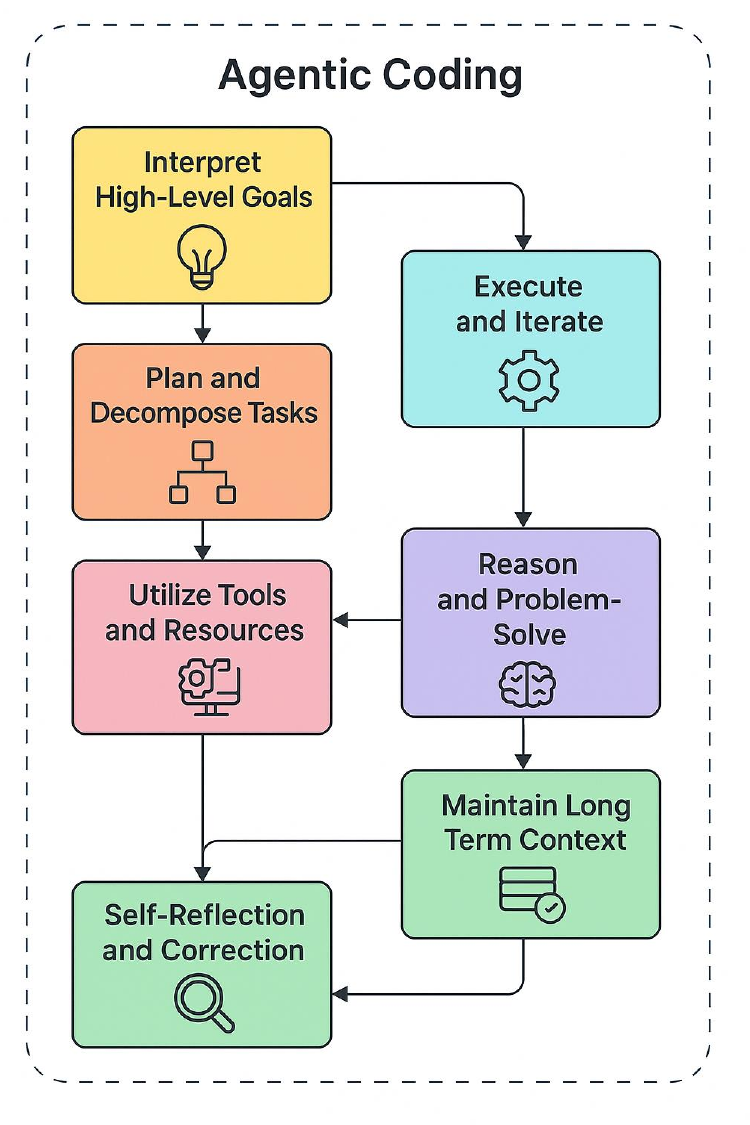}
    \caption{\textbf{Core Capabilities in Agentic Coding:} Illustrating the sequential and interconnected capabilities of agentic coding: Interpret High-Level Goals, Plan and Decompose Tasks, Utilize Tools and Resources, Execute and Iterate, Reason and Problem-Solve, Maintain Long-Term Context, and Self-Reflection and Correction within autonomous software agents}
    \label{fig:agenticcoding}
\end{figure}

Agentic coding as illustrated in Figure \ref{fig:agenticcoding} represents a paradigmatic shift in AI-assisted software engineering. Unlike vibe coding, where LLMs operate as conversational co-pilots \cite{wijaya2025readme, maes2025gotchas}, agentic coding systems delegate substantial cognitive and operational responsibility to autonomous or semi-autonomous software agents. These agents are capable of planning, executing, and verifying complex software tasks transforming natural language instructions into robust, testable code with minimal human guidance \cite{kumar2025tfhe, roychoudhury2025agentic}. Architecturally, this requires the convergence of goal planning, task decomposition, execution environments, safety infrastructure, and continuous feedback mechanisms \cite{zhang2025adaptive, kumar2025tfhe}.

The core philosophy of agentic coding is delegated autonomy. Developers specify high-level objectives such as “integrate an external API,” “refactor backend routing,” or “set up CI workflows,” while the agent assumes responsibility for determining and executing the steps needed to accomplish those goals. This transforms the human’s role from low-level implementer to a system-level supervisor and goal-setter.

Agentic agents exhibit the following core capabilities:

\begin{itemize}
    \item \textbf{Interpret High-Level Goals:} Agentic systems parse natural language prompts that span multiple files, layers, or components \cite{sapkota2025ai}. For instance, Jules (developed by Google) \href{https://jules.google/}{Jules Link} can respond to queries such as “integrate the Google Gemini API into the R1 robot” by identifying relevant entry points in the codebase.
    
    \item \textbf{Plan and Decompose Tasks:} Upon receiving a request, agents create internal execution plans. Jules \href{https://jules.google/}{Jules, Google}, for example, breaks down the task into subtasks such as API research, data structure design, code insertion, documentation updates, and test plan execution.
    
    \item \textbf{Utilize Tools and Resources:} Agents autonomously interact with file systems, compilers, interpreters, test suites, Git repositories, APIs, and even browsers. In Codex \href{https://openai.com/codex/}{Codex, OpenAI}, sandboxed environments are spun up for each task, with independent dependencies and runtime isolation.
    
    \item \textbf{Execute and Iterate:} Agents can modify source code (e.g., changing `RoboLogic.cs`), test their output, log failures, and retry iteratively. Codex \href{https://openai.com/codex/}{Codex, OpenAI}, for instance, can automatically run `git diff`, apply patches, and generate pull requests.
    
    \item \textbf{Reason and Problem-Solve:} When encountering edge cases, agents apply heuristics, run static analysis, or search documentation. In Jules’s integration task, error handling included adjusting response parsers and dynamically reconfiguring the Unity Inspector \href{https://jules.google/}{Jules, Google}.
    
    \item \textbf{Maintain Long-Term Context:} Codex maintains session state over complex multi-step tasks, managing API keys, dependencies, and environment variables \href{https://openai.com/codex/}{Codex, OpenAI}. Persistent memory and vector store integration enable agents to reference earlier instructions and code changes.
    
    \item \textbf{Self-Reflection and Correction:} Emerging systems implement internal evaluation. Agents like Codex log their decision trees, summarize actions, propose revisions, and retry failed steps autonomously presenting diffs and execution summaries to the user \href{https://openai.com/codex/}{Codex, OpenAI}.
\end{itemize}

\begin{figure}
    \centering
    \includegraphics[width=\linewidth]{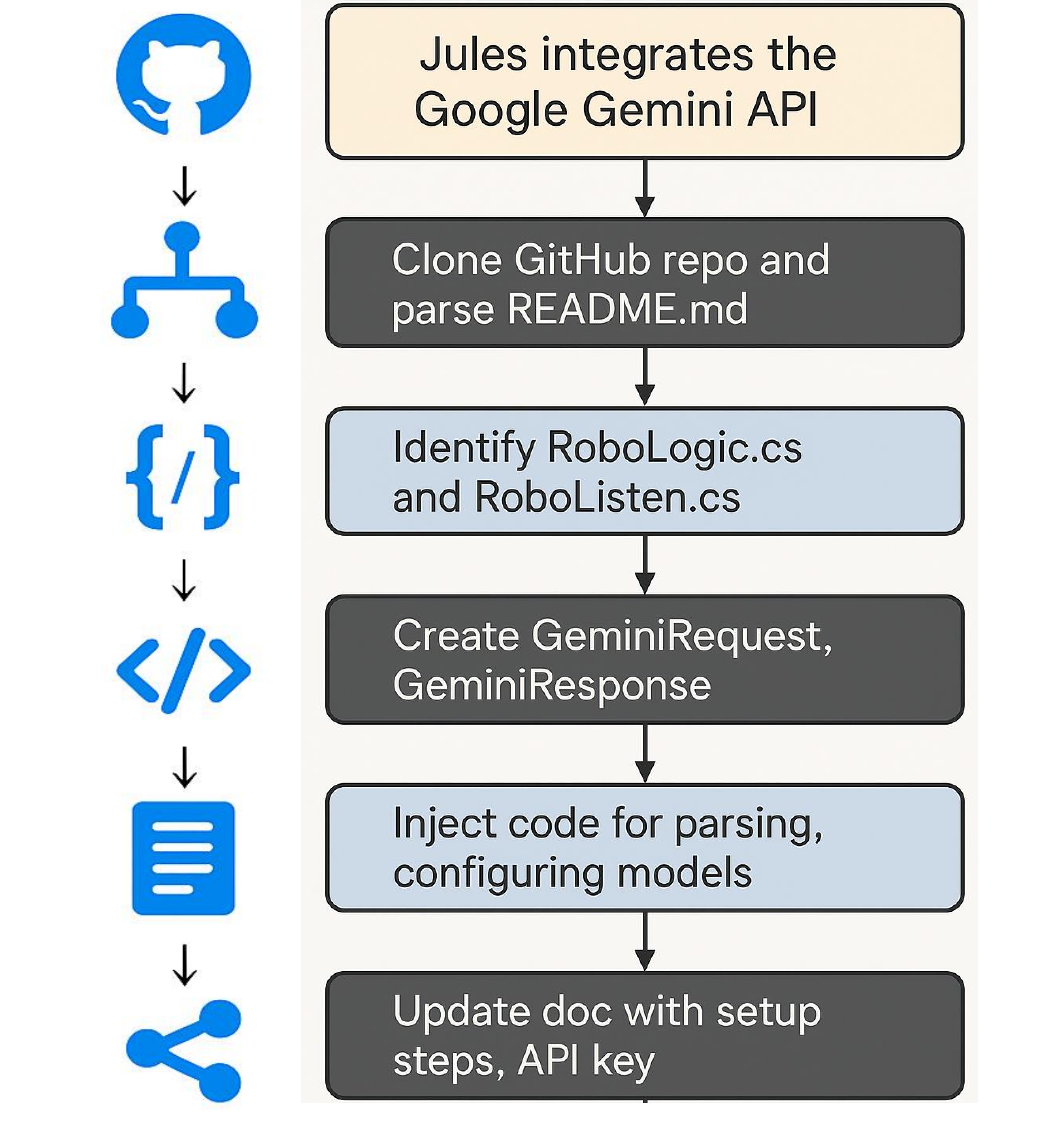}
    \caption{Agentic Coding Example-API Integration with Jules. This flowchart illustrates how Jules autonomously integrates the Google Gemini API, executing tasks from repository analysis and code modification to documentation and Git operations demonstrating a complete, multi-step software development workflow typical of advanced agentic coding systems.}
    \label{fig:ExammleJule}
\end{figure}

The human-agent interaction remains iterative but high-level. In Jules, for instance, developers are presented with reviewable summaries (“Ready for Review”) and given options to approve, revise, or publish branches. In Codex, task outcomes are presented with logs, diffs, and test results for validation before pushing to GitHub. When instructed to integrate the Google Gemini API into a robotics codebase, the agentic coding system Jules demonstrated a multi-step, autonomous workflow that exemplifies the principles of agentic software development. As illustrated in Figure~\ref{fig:ExammleJule}, Jules began by cloning the target GitHub repository and analyzing the README.md file to establish project context and configuration. It then autonomously identified relevant integration points namely RoboLogic.cs and RoboListen.cs as the scripts most suitable for modification. The agent proceeded to generate two new data classes, GeminiRequest and GeminiResponse, to support the structure of the API's request/response handling. It injected the necessary code to parse responses from the Gemini API and configured model parameters to be adjustable via Unity Inspector fields, streamlining developer interaction with the AI integration. To ensure usability and reproducibility, Jules updated the documentation, outlining API key requirements and configuration steps. Finally, it committed all modifications to a newly created Git branch and presented the changes for review. This sequence not only reflects an end-to-end software modification task performed autonomously but also highlights the value of agentic systems in managing complex API integrations, combining planning, reasoning, documentation, and version control in a unified pipeline.

Table \ref{tab:vibe_vs_agentic_taxonomy} provides a structured taxonomy comparing the core characteristics, execution roles, and interaction patterns of Vibe Coding and Agentic Coding. It highlights how these paradigms differ in autonomy, developer responsibility, tool integration, and system maturity, offering a comprehensive view of their conceptual and technical distinctions.

\vspace{0.5 cm}

\begin{table*}[ht!]
\centering
\caption{Comparative Taxonomy of Vibe Coding versus Agentic Coding Paradigms}
\label{tab:vibe_vs_agentic_taxonomy}
\begin{tabular}{
  >{\RaggedRight\arraybackslash}p{3.5cm} 
  >{\RaggedRight\arraybackslash}p{6.5cm} 
  >{\RaggedRight\arraybackslash}p{6.5cm} 
}
\toprule
\textbf{Aspect} & \textbf{Vibe Coding} & \textbf{Agentic Coding} \\
\midrule
\textbf{Core Philosophy} & 
Human-AI co-creation and intuitive partnership. Developer guides AI with high-level "vibes" and iterative feedback \cite{lou2025unraveling}. Focus on augmenting individual developer productivity \cite{shen2025ideationweb, wang2024rocks}. & 
AI as an autonomous or semi-autonomous task executor. Developer delegates complex goals, and AI plans and executes with higher-level human oversight \cite{hammond2025multi, kasirzadeh2025characterizing}. Focus on automating larger development tasks \cite{tufano2024autodev, sager2025ai}. \\
\addlinespace 
\textbf{AI Autonomy Level} & 
Low to Moderate: Acts as a responsive assistant, executing specific, developer-defined tasks \cite{wen2024autodroid, xu2025every} (e.g., "write this function," "refactor this class"). Requires continuous human prompting for sequential steps \cite{ma2025should, wu2022promptchainer}. & 
Moderate to High: Acts as a proactive executor \cite{qian2023communicative, triedman2025multi}. Can independently plan, decompose, and execute a series of sub-tasks based on high-level goals with significantly less step-by-step human intervention. Can make intermediate decisions \cite{sapkota2025ai}.\\
\addlinespace
\textbf{Developer's Primary Role} & 
Director, Co-Pilot, Prompter: Actively involved in defining micro-tasks, prompting for code, reviewing each generated piece, debugging details, and integrating components \cite{luo2025large}. & 
Architect, Project Manager, Supervisor: Defines high-level strategic goals, system architecture, constraints. Monitors agent's progress, provides strategic guidance, and validates outcomes \cite{krishnan2025advancing, habler2025building}. \\
\addlinespace
\textbf{Interaction Model} & 
Highly conversational and iterative prompt-response cycles \cite{rath2025structured}. Short, frequent feedback loops \cite{dai2023llm, kazemitabaar2023novices}. Developer continuously refines prompts or code \cite{white2024chatgpt}. & 
Goal delegation and monitoring. Human sets a complex goal; agent plans, executes, and provides periodic updates or requests high-level clarification \cite{luo2025large}. Longer, more complex execution cycles with less frequent, but more strategic, human check-ins \cite{wang2025training}. \\
\addlinespace
\textbf{Planning \& Task Decomposition} & 
Primarily human-led. Developer breaks down features into smaller, promptable tasks for the AI. AI's planning is mostly local to the current prompt \cite{ding2024reasoning}. & 
Primarily AI-led within defined constraints. The agent autonomously decomposes high-level goals into an actionable sequence of sub-tasks and dependencies \cite{ferrag2025llm, sapkota2025ai}. \\
\addlinespace
\textbf{Tool Utilization by AI} & 
Primarily code generation. Use of other development tools (compilers, Git, linters) is typically initiated and managed by the human based on AI's code output \cite{lyu2024automatic}. & 
Can autonomously interact with and orchestrate a suite of development tools (e.g., compilers, interpreters, version control systems, package managers, APIs, web browsers for research/testing) as part of its execution plan \cite{borghoff2025human, pan2025codecor}. \\
\addlinespace
\textbf{Execution Environment} & 
Often generates code for execution in an external IDE or a minimally integrated environment \cite{li2024enhancing}. The developer is responsible for setting up and managing the runtime \cite{sun2024llm}. & 
Typically requires integrated or sandboxed execution environments with robust logging, resource management, and security guardrails to safely execute agent-generated code and tool commands \cite{guo2024redcode, qiang2025mle}. \\
\addlinespace
\textbf{Error Detection \& Handling} & 
Errors are often identified by the developer post-generation during manual code review, testing, or via IDE linters/compilers \cite{tufano2024autodev}. AI assists in fixing errors when prompted with specific error context \cite{poitras2024generative, kazemitabaar2023novices}. & 
Aims for more autonomous error detection by the agent (e.g., through automated test execution, static analysis, or internal consistency checks) \cite{nunez2024autosafecoder} and may attempt self-correction, with human intervention required for complex or novel failures \cite{song2025progco, askari2025magic}. \\
\addlinespace
\textbf{Typical Task Scope} & 
Component-level tasks: generating functions, UI elements, simple scripts, unit tests for specific functions, drafting documentation, refactoring small code sections \cite{wei2024requirements}. & 
Feature-level or even simple application-level tasks: end-to-end feature implementation (e.g., user authentication system), complex multi-file refactoring, system migrations, comprehensive test suite generation, CI/CD pipeline automation \cite{fang2025mlzero, agashe2024agent, tzanis2025maistro}. \\
\addlinespace
\textbf{Context Management by AI} & 
Primarily relies on the developer to provide and maintain context within a session, often limited by the LLM's inherent context window \cite{mohsin2024can, nam2024using}. Human bridges context across longer interactions. & 
Designed for more sophisticated and persistent context management, enabling reasoning over more extensive project information \cite{tolzin2025uncovering}, documentation, and past interactions across longer, multi-step tasks \cite{wang2024rocks, shentu2024llms}. \\
\addlinespace
\textbf{Primary Output Focus} & 
Generation of code snippets, functions, or small modules that the developer then integrates and refines \cite{joel2024survey, wei2024requirements, nam2024using}. & 
Generation of more complete, interconnected systems or features \cite{wu2023autogen, ishibashi2024self, luo2025large}, potentially including code, tests, configurations, and documentation, requiring less manual integration by the developer \cite{rasheed2024codepori, puvvadi2025coding, tang2025metachain}. \\

\addlinespace
\textbf{Maturity and Current Availability} & 
Widely adopted in tools like GitHub Copilot \href{https://github.com/features/copilot}{Link}, Cursor \href{https://www.cursor.com/}{Link}, and Replit Agent \href{https://replit.com/}{Link}. & 
Emerging and experimental; seen in platforms like AutoGen \href{https://microsoft.github.io/autogen/stable//index.html}{ Microsoft AutoGen}, CrewAI \href{https://www.crewai.com/}{Crew AI link}, Codex by OpenAI \href{https://chatgpt.com/codex}{Codex Link} LangChain Agents \href{https://python.langchain.com/docs/introduction/}{LangChain}, and early products such as Devin \href{https://devin.ai/}{Devin AI} \\

\bottomrule
\end{tabular}
\end{table*}

\subsubsection{Conceptual Architecture of Agentic Systems}

Agentic coding systems (Figure \ref{fig:agenticcoding})  are architecturally distinct from prompt-driven LLM tools, exhibiting a modular and cognitively looped design tailored for autonomous software engineering. At their core, agentic platforms such as Codex and Jules integrate planning, execution, tool interaction, and evaluation into a cohesive, goal-driven framework.

The conceptual architecture typically comprises several interlinked components. A \textit{core reasoning engine}, powered by a LLM, interprets high-level developer instructions and generates actionable plans. This is supported by a \textit{planning module}, which decomposes abstract goals into a sequence of structured sub-tasks, using mechanisms such as chain-of-thought prompting or hierarchical task networks \cite{liu2025nature, sapkota2025ai, schneider2025generative}. To enable environmental interaction, a \textit{tool use module} grants agents the ability to execute commands or access APIs via function calling \cite{bornet2025agentic, mohammadi2025pel}. This includes capabilities such as modifying configuration files, running shell commands, or interacting with Git repositories \cite{liu2025advances}.

A critical feature is the presence of \textit{memory and context management} \cite{xu2025mem, miehling2025agentic}, facilitating persistent state tracking across multi-step workflows. Agents leverage both short-term working memory and long-term retrieval-augmented memory to maintain coherence in extended tasks. All actions are executed within an \textit{isolated sandboxed environment} that enforces system safety through resource constraints, permission scoping, and rollback mechanisms \cite{sapkota2025ai}.

Feedback is central to the agentic paradigm where agents incorporate results from automated tests, logs, or human feedback through \textit{evaluation and learning mechanisms}, adjusting future behavior accordingly \cite{fabian2025agentic, kamalov2025evolution}. Architectures may further include an \textit{orchestration layer} that coordinates specialized sub-agents (e.g., planner, coder, tester, documenter), facilitating parallelism and modular division of labor \cite{liu2025advances}.

As illustrated in systems like Codex  \href{https://openai.com/codex/}{Codex, OpenAI}, this architecture transitions AI from a passive tool to an active collaborator capable of self-directed planning, decision-making, and refinement. These agents operate not merely as extensions of developer intent but as semi-autonomous entities capable of transforming high-level specifications into verifiable software artifacts \cite{worring2025multimedia, feng2025integration}. Agentic coding thus lays the foundation for scalable, adaptable, and increasingly intelligent development pipelines in real-world programming ecosystems \cite{goh2025development, kshetri2025transforming}.

\subsubsection{Shift in Developer Interaction and Control}

The interaction paradigm in agentic coding represents a fundamental departure from the co-piloting model of vibe coding \cite{patil2024towards, granberry2025seeking}. Rather than engaging in direct, iterative instruction at the function or line level, the developer assumes a supervisory role defining the mission, monitoring system behavior, and validating outcomes \cite{haque2025llms}. This transition from procedural engagement to goal-level delegation reflects a broader cognitive and operational realignment in human-AI collaboration.

At the outset, the developer is responsible for \textit{mission specification}, articulating high-level objectives, architectural constraints, and system-level requirements \cite{yu2025multi, deng2025llm}. These inputs may encompass functional targets (e.g., “integrate external API for user analytics”), non-functional constraints (e.g., security, latency, portability), or domain-specific standards. The agent then plans and initiates the execution process autonomously \cite{zhang2025webpilot, hexmoor2025behaviour, wan2025reca}.

Throughout execution, the developer assumes the role of an \textit{observer and strategic guide}, reviewing real-time logs, intermediate artifacts, and agent-generated plans \cite{chakraborty2025collab, moshkovich2025beyond}. This includes evaluating execution traces, test results, and change diffs \cite{robeyns2025self, gandhi2025researchcodeagent}. Intervention may be necessary when the agent encounters ambiguous requirements, edge cases beyond its training distribution, or tasks that involve ethical, legal, or architectural judgment. Critically, the developer also acts as the \textit{final verifier} \cite{nadimi2025verimind, sung2025verila, qi2025agentif}. Before any integration or deployment, the human evaluates the full solution ensuring correctness, compliance, and alignment with project vision \cite{wang2025comprehensive}. This oversight transforms the developer’s responsibilities from tactical implementation to strategic assurance and decision validation.

This evolving model requires a distinct set of cognitive and technical competencies. Developers must develop fluency in “agent management” understanding agent capabilities, interpreting failure modes, designing effective prompts and constraints, and deploying diagnostic tools when agents deviate from expected behavior \cite{liu2025advances}. The trust placed in the agent must be balanced with a readiness to intercede, especially in high-risk or safety-critical contexts. Ultimately, the agentic interaction model foregrounds the human as a system architect, supervisor, and ethical gatekeeper, overseeing a semi-autonomous AI collaborator \cite{sapkota2025ai, liu2025advances}. This shift not only augments developer productivity but also redefines the nature of software engineering in AI-mediated environments. Additional detailed distinctions between Vibe Coding and Agentic Coding including differences in autonomy, task scope, error handling, and developer role are comprehensively summarized in Table~\ref{tab:vibe_vs_agentic_taxonomy}.

\begin{figure*}[ht!]
    \centering
    \includegraphics[width=\linewidth]{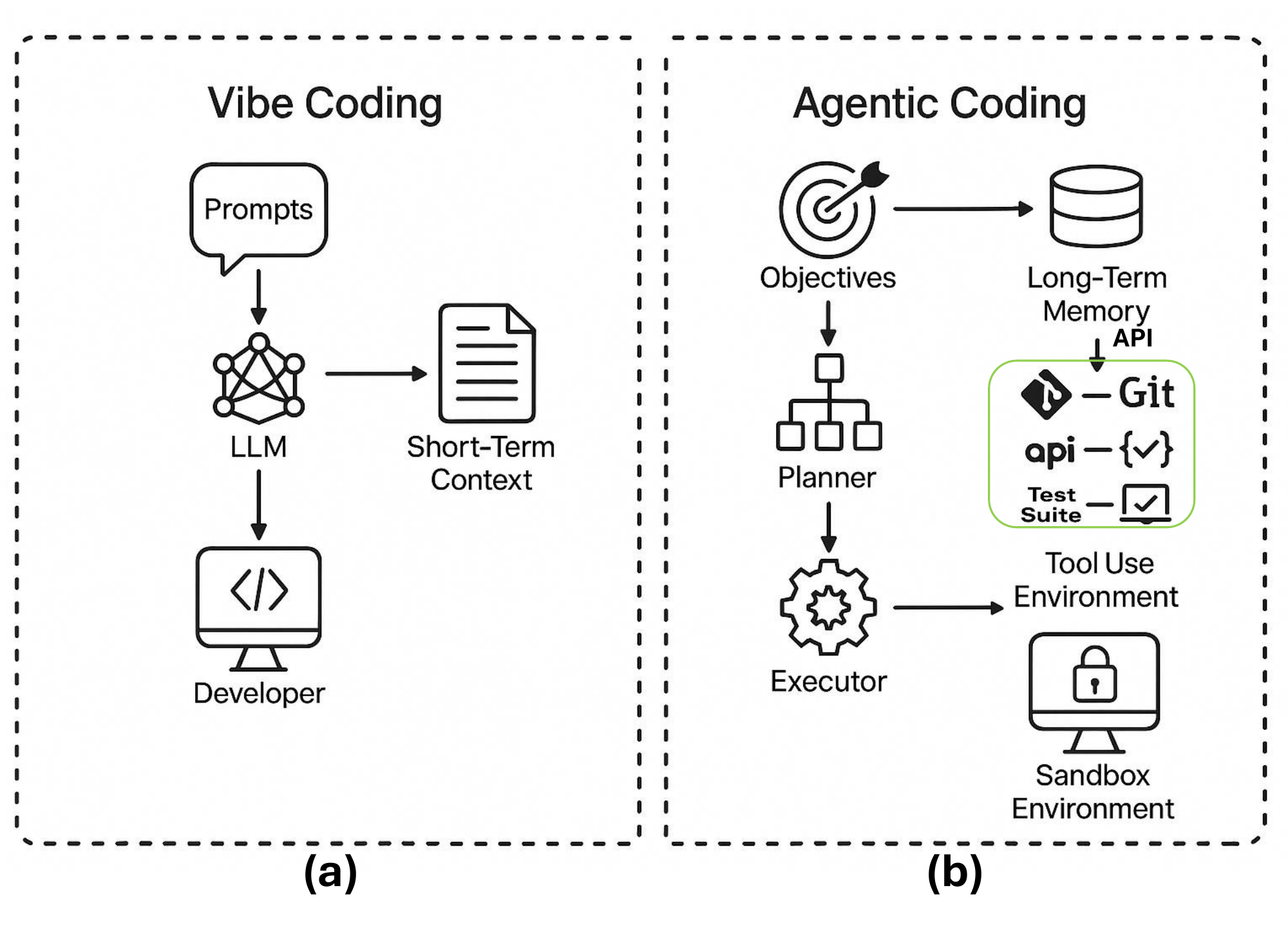}
    \caption{\textbf{Comparative Architecture of Vibe Coding and Agentic Coding} (a) Vibe Coding: Developers provide prompts to an LLM within an IDE or web interface. The workflow relies on short-term context and manual execution, testing, and integration. (b) Agentic Coding: Developers define objectives processed by a planner, long-term memory, and executor modules. Agents autonomously use tools within sandboxed environments to complete multi-step workflows.}
    \label{fig:architectural}
\end{figure*}
\section{Technical Architecture and Capabilities}
Although both vibe coding and agentic coding harness LLMs to augment software development as depicted in Figure \ref{fig:architectural}, their architectural intent and implementation are fundamentally distinct. Vibe Coding (Figure \ref{fig:architectural}a)  operates through developer-initiated, prompt-based interactions within IDEs or web-based environments, emphasizing conversational co-creation and low-friction prototyping. In contrast, Agentic Coding (Figure \ref{fig:architectural}b)  is grounded in delegated autonomy: developers specify high-level objectives, and intelligent agents often composed of planner, executor, and toolchain modules carry out multi-step coding workflows, potentially invoking compilers, APIs, test runners, and version control systems without continuous human supervision. To articulate these differences, this section presents a detailed architectural analysis through layered diagrams, pseudocode abstractions, and systematic tabular comparisons. The core architectural contrasts ranging from context management and multi-agent orchestration to execution sandboxing and CI/CD integration are summarized in Table~\ref{tab:architecture_comparison}, offering researchers and system designers a clear framework for understanding the capabilities and trade-offs of each model. Additionally, we explore how feedback loops, validation protocols, and tool autonomy shape each paradigm’s suitability for different use cases, from rapid prototyping to enterprise-scale automation. By formalizing these architectural features, this section contributes a foundational taxonomy for evaluating emerging AI coding frameworks, informing both engineering decisions and future research in agentic software systems.

\begin{table*}[ht!]
\centering
\caption{Architectural Comparison Between Vibe Coding and Agentic Coding Systems}
\label{tab:architecture_comparison}
\begin{tabular}{
  >{\RaggedRight\arraybackslash}p{3.8cm}
  >{\RaggedRight\arraybackslash}p{6.1cm}
  >{\RaggedRight\arraybackslash}p{6.1cm}
}
\toprule
\textbf{Architectural Component} & \textbf{Vibe Coding} & \textbf{Agentic Coding} \\
\midrule
\textbf{System Orchestration} &
Typically centralized around IDE-based LLM plugins or browser UIs. Multi-agent layers are lightweight and session-bound. Human initiates each cycle. &
Hierarchical agent controllers with planner–executor modules. Autonomy allows task spawning, recursive loops, and tool orchestration without new human prompts. \\
\addlinespace
\textbf{Prompt Handling Layer} &
Interactive, session-limited prompting with optional pre-prompts or in-context examples. Managed by user interface or IDE plugin. &
Semantic goal parsing with persistent memory modules, chaining of system, user, and tool prompts. Task graphs and agent routing handled internally. \\
\addlinespace
\textbf{Memory and Context Architecture} &
Short-term context via prompt window + optionally dynamic context via RAG (Retrieval-Augmented Generation). Persistent memory rare. &
Uses vector memory stores, session-level caching, external knowledge bases, and context prioritization queues for long-horizon state awareness. \\
\addlinespace
\textbf{Execution Engine} &
Human-managed execution loop; AI produces artifacts for IDEs, terminals, or hosted sandboxes. Limited or no write-access to environment. &
Embedded, sandboxed, or containerized runtime. Agents execute, modify, and validate code via system calls, test runners, compilers, or browsers. \\
\addlinespace
\textbf{Multi-Agent Coordination} &
Optional, and usually predefined (e.g., scaffolder + code generator). Agents do not independently collaborate or plan. &
Core architectural feature. Specialized sub-agents (e.g., coder, tester, reviewer, fixer) collaborate, pass state, and report to a planner agent. \\
\addlinespace
\textbf{Toolchain Integration} &
Limited; human bridges between AI output and tools like Git, CI/CD, or package managers. Tool actions are suggested, not executed. &
Full-stack tool orchestration (e.g., Git commits, API calls, dependency management, monitoring). Agents autonomously trigger workflows. \\
\addlinespace
\textbf{Validation Pipeline} &
Linting, test generation, and review are human-invoked or post-hoc. AI may help create tests, but does not own testing pipeline. &
Integrated QA loop with automatic test synthesis, execution, and patching. Agents evaluate outputs, revise plans, and regenerate on failure. \\
\addlinespace
\textbf{Security and Guardrails} &
Basic static analysis or organizational linters applied post-generation. Security remains human-enforced. &
Embedded static and dynamic scanners, policy-as-code integration, and gated deploy checkpoints. Security policies enforced during execution. \\
\addlinespace
\textbf{Observability and Feedback} &
Lightweight usage metrics via IDEs or logs. Feedback loop is implicit (developer adjusts prompt) and local. &
Telemetry-driven feedback to prompt-engineering layer. Continuous refinement via logs, metrics, exceptions, and outcome evaluations. \\
\addlinespace
\textbf{Deployment and CI/CD} &
Output consumed and deployed by human; CI/CD setup is manual or external. &
Agents may set up and trigger CI/CD workflows, generate configs (e.g., GitHub Actions), and validate deployment success or rollback. \\
\bottomrule
\end{tabular}
\end{table*}

\subsection{Execution Models: Comparative Analysis of Architectural Design}

\subsubsection{Vibe Coding Interfaces and Developer-Driven Execution}

Vibe coding architectures operate primarily through lightweight, stateless interfaces where LLMs serve as code-generation engines embedded in developer-centric environments such as IDEs, browser-based editors (e.g., Replit \href{https://replit.com/}{Replit AI}), or terminal integrations \cite{zhang2024codeagent, nunez2024autosafecoder}. The execution model is explicitly decoupled from the generation pipeline LLMs suggest or write code in response to high-level prompts, but the responsibility for integration, execution, testing, and debugging remains with the human developer \cite{jiang2024survey, chen2024survey, rahman2025marco}. The developer copies generated snippets into their runtime environment, configures test cases, and manually interprets any resulting behavior.

This model emphasizes flexibility and creativity during early-stage development or rapid prototyping, leveraging prompt-response cycles to accelerate code synthesis. However, from an architectural standpoint, it exhibits a \textit{passive execution pipeline}. There is no embedded runtime or agent-native validation loop \cite{wang2025internet, xu2025forewarned}. Instead, testing and validation are handled through external services unit test frameworks, CI/CD tools, or manual test execution within local or cloud IDEs \cite{kothamali2025ai}.

This asynchronous, generation-first design allows LLMs to focus on semantic synthesis and reuse of learned patterns \cite{pu2025assistance}, but introduces latency in feedback loops and a higher cognitive burden on the developer \cite{zamfirescu2025beyond, gu2025challenges}. The architecture lacks internal state management, agent memory, or runtime enforcement, reflecting its reliance on human-driven control over execution and validation.

\subsubsection{Agentic Coding Architectures and Autonomous Execution Pipelines}

Compared to vibe coding, agentic coding systems incorporate fully integrated execution pipelines as a first-class architectural feature \cite{walters2025eliza, berberi2025machine}. These systems embed containerized, policy-constrained runtime environments such as Docker instances \cite{hu2025llm, chauhan2025llm, wolflein2025llm}, WASM runtimes \cite{tang2025metachain, ye2025maslab}, or lightweight QEMU-based emulators directly into the development agent’s operational core \cite{kanellopoulos2025virtuoso}. Within these sandboxes, autonomous agents can not only generate code but also execute, test, and iteratively refine it without requiring human intervention for each step \cite{pan2025codecor, almorsi2025guided}.

Agentic execution architectures are characterized by modular task graphs where planner components decompose user goals into executable sub-tasks, and executor agents interact with the runtime to carry them out \cite{miloradovic2023optimizing}. This allows for a tight coupling between generation, execution, and feedback \cite{wang2024executable}. Agents dynamically manage system state, interact with file systems \cite{cao2024managing}, perform queries \cite{mei2024aios,li2024personal}, analyze logs \cite{ferrag2025llm}, and retry failed attempts based on real-time results \cite{liu2023llm, roy2024exploring}. Security and control are maintained through fine-grained resource isolation sandboxing policies govern memory usage, file I/O, and network access \cite{luo2025large, long2025survey}.

\begin{tcolorbox}[colback=blue!5!white, colframe=blue!75!black, title=Agentic Coding Workflow]
\begin{itemize}
    \item Developer prompt: ``Optimize SQL joins for user reporting.''
    \item Agent loads sandbox environment.
    \item Analyzes ORM model structure.
    \item Refactors queries and validates execution plan.
    \item Deploys tested code to staging environment.
\end{itemize}
\end{tcolorbox}

This closed-loop, self-evaluating architecture reduces reliance on the human as a runtime operator and increases system autonomy. It supports advanced use cases such as multi-file refactoring, regression analysis, and continuous integration with minimal human oversight. Architecturally, this marks a transition from \textit{interactive co-programming} to \textit{autonomous software engineering}, where execution is proactive, contextual, and adaptively managed by intelligent agents.

\subsection{Autonomy and Feedback Loops in Vibe and Agentic Coding Paradigms}

\subsubsection{Vibe Coding: Human-Centric Control and Reactive Feedback}

Vibe coding architectures operate under a fundamentally reactive model, wherein the human developer remains the sole agent responsible for validation, error detection, and iterative refinement \cite{wang2024rocks}. The LLM acts as a stateless code synthesis engine generating outputs in response to prompt instructions but without any intrinsic feedback mechanism or self-evaluation capacity \cite{agashe2024agent}. As such, the feedback loop exists entirely outside the system and is mediated by the developer through post-hoc testing, debugging, and prompt refinement \cite{xie2024waitgpt, ehsani2025towards}.

This model affords significant flexibility in exploratory or creative coding sessions. Developers may use short, expressive prompts (e.g., ``Add JWT authentication to the login flow'') and immediately evaluate the output in their IDE or test environment. However, when prompts are vague or underspecified (e.g., ``Make this more secure''), the LLM's lack of situational awareness and task-level memory often leads to hallucinated or ambiguous outputs.

\begin{tcolorbox}[colback=yellow!10!white, colframe=orange!70!black, title=Prompt Engineering Insight]
\textbf{Specific prompt:} ``Add role-based access control using JWTs and restrict admin endpoints.'' \\
\textbf{Vague prompt:} ``Make this more secure.'' \\
\textbf{Outcome:} The former yields focused middleware code with user roles; the latter returns generic suggestions like hashing passwords twice or limiting requests, often misaligned with project context.
\end{tcolorbox}

Due to its lack of autonomous validation \cite{alami2025accountability}, vibe coding systems are limited in production environments where reliability, regression testing, and integration constraints are critical \cite{fakhoury2024llm, lyu2024automatic}. Developers must manually run tests, validate results, and reframe prompts for each iteration, rendering the process iterative but human-dependent. While suitable for front-end prototyping, documentation drafting, or low-risk automation, the absence of self-driven error correction limits its robustness in complex systems.

\subsubsection{Agentic Coding: Goal-Driven Autonomy with Feedback-Integrated Execution}
In contrast, agentic coding frameworks are designed with feedback-driven autonomy as a core architectural principle \cite{huang2025towards}. Agents operate through multi-level feedback loops that include planning, execution, testing, evaluation, and corrective iteration all orchestrated without human prompting between steps \cite{  sapkota2025ai}. This architecture draws from reinforcement learning, symbolic planning, and black-box evaluation strategies to enable continuous improvement within a coding session.

A typical agentic workflow begins with a high-level task objective (e.g., ``Build a PostgreSQL-backed user analytics dashboard''), which is decomposed into subtasks using internal planning modules. Each subtask (e.g., schema generation, query writing, UI wiring) is independently implemented and validated via in-agent execution environments. Failures trigger internal debugging logic, resulting in retrials, log inspection, or substitution strategies.

\begin{tcolorbox}[title=Agent Feedback Algorithm, colback=white, colframe=black]
\begin{verbatim}
Input: Task Objective (T)
Output: Verified Implementation (I)

Decompose(T) -> [t1, t2, ..., tn]
For each ti:
    Implement(ti)
    Run Test(ti)
    if fail: Debug(ti), Repeat
Aggregate([t1...tn]) -> I
Return I
\end{verbatim}
\end{tcolorbox}

This closed-loop feedback enables high fidelity in repetitive and deterministic programming contexts \cite{liu2025advances, sapkota2025ai}, such as dependency management, CI/CD configuration, or auto-generating test suites for large-scale systems. For example, an agent tasked with ``Migrate project from JavaScript to TypeScript'' will iterate through module identification, static analysis, AST rewriting, and runtime testing without developer intervention at each step.

Unlike vibe systems, agentic architectures support telemetry, traceability, and performance metrics at each layer, enabling outcome-aware re-planning and model fine-tuning. The result is an execution pipeline that resembles autonomous software engineering rather than assisted coding capable of aligning long-term goals with tactical implementation across multiple files, systems, and APIs.

\subsection{Safety, Explainability, and System Constraints}

\subsubsection{Vibe Coding: Limited Guardrails and Post-Hoc Safety Mitigation}

Vibe coding environments, by design, prioritize fluidity of interaction and developer creativity over integrated safety controls. The underlying architecture does not include runtime enforcement mechanisms, making safety and explainability externalized concerns. Outputs are typically generated without runtime awareness, leading to several risks in security-sensitive or regulated environments.

A critical architectural limitation is the absence of \textit{execution traceability}. Since LLMs are stateless within a session, they cannot record, annotate, or justify their decisions unless explicitly prompted to do so \cite{north2024code, hassine2024llm, chew2023llm}. This lack of interpretability becomes particularly concerning when the AI injects code with hardcoded credentials, insecure API calls \cite{mohsin2024can, mousavi2024investigation}, or unsafe permission scopes problems often observed in rapid prototyping workflows \cite{guo2024redcode}.

\begin{tcolorbox}[colback=red!5!white, colframe=red!75!black, title=Common Vibe Coding Risks]
\begin{itemize}
    \item \textbf{Hardcoded secrets:} Generated code may embed plaintext API keys or passwords.
    \item \textbf{Insecure defaults:} Lack of input sanitization, overly permissive CORS headers.
    \item \textbf{No audit trails:} Developer cannot inspect prior outputs unless manually documented.
\end{itemize}
\end{tcolorbox}

To mitigate these risks, developers often rely on external static analysis tools e.g., SonarQube \href{https://www.sonarsource.com/products/sonarqube/}{Sonar}, CodeQL \href{https://codeql.github.com/}{Github link}, or ESLint security plugins \href{https://eslint.org/}{ESList} to perform post-generation audits. These tools can flag anti-patterns, insecure imports, or style violations. However, these solutions operate independently of the LLM and require the developer to integrate them manually into their pipeline. As a result, the responsibility for enforcing safety, explainability, and governance in Vibe Coding rests solely on the human-in-the-loop, limiting its applicability in high-assurance domains like finance, healthcare, or enterprise DevOps.

\subsubsection{Agentic Coding: Embedded Safeguards and Transparent Execution}

Agentic coding frameworks are designed with embedded safety constraints \cite{yang2024plug, he2024frontiers}, explainability mechanisms \cite{sanwal2025layered, sapkota2025ai}, and runtime isolation policies \cite{wu2024isolategpt, ashrafi2025enhancing, triedman2025multi}. These systems are designed to emulate production-grade deployment scenarios in microcosm allowing agents to safely execute, debug, and iterate while maintaining verifiable compliance with security and governance policies \cite{de2025open, syros2025saga}.

The first tier of architectural safeguards involves resource and namespace isolation. Agent containers run within sandboxed environments where access to file systems, memory, CPU, and network interfaces is tightly scoped and rate-limited \cite{hu2025llm, tang2025metachain}. For example, an agent modifying YAML configuration files may only access a whitelisted directory tree, preventing accidental file system corruption or privilege escalation.

\begin{tcolorbox}[colback=blue!5!white, colframe=blue!75!black, title=Agentic Safety Features]
\begin{itemize}
    \item \textbf{Namespace isolation:} Prevents unauthorized file system access.
    \item \textbf{Resource limits:} Controls execution via CPU/memory quotas.
    \item \textbf{Logging hooks:} Captures all agent actions, prompts, and test results.
    \item \textbf{Rollback triggers:} Reverts files or state if tests fail or errors occur.
\end{itemize}
\end{tcolorbox}

Explainability is built into the execution graph \cite{ferrag2025llm, sapkota2025ai}. Tools like Claude Code \href{https://www.anthropic.com/claude-code}{Claude Code Link}, Amazon Q Developer \href{https://aws.amazon.com/q/developer/}{Amazon-Q-Developer Link}, and Devika \href{https://devikaai.org/}{Devika AI} log every decision node and code transformation, enabling post-hoc inspection and diff analysis. These logs not only serve as audit trails for compliance but also allow developers to interpret the agent's reasoning chain for example, why it refactored a function, replaced a package, or reordered a CI pipeline.

Such mechanisms elevate agentic systems from mere automation engines to auditable \cite{barua2024exploring, tang2025metachain}, controlled execution environments \cite{wang2024executable}. Furthermore, the rollback infrastructure ensures that the system can revert unintended side effects, thereby reducing the risk of silent failures or irreversible changes \cite{zhang2024breaking, cemri2025multi}. These features make agentic coding architectures more aligned with enterprise-grade reliability and explainability standards, distinguishing them as preferable frameworks for autonomous software engineering in safety-critical domains.

\textbf{Overall Architectural Comparison and Illustrative Application}

To synthesize the architectural contrasts between vibe coding and agentic coding, we present a comparative analysis focused on execution fidelity, safety, and autonomy. Table~\ref{tab:execution_safety_comparison} summarizes key differences in system capabilities across multiple operational dimensions. Notably, agentic systems are characterized by in-sandbox execution, embedded validation, and robust safety mechanisms \cite{sapkota2025ai}, whereas vibe coding tools operate in a generation-first manner, relying on human oversight for execution, testing, and risk mitigation \cite{rao2025navigating}.

\begin{table}[ht!]
\centering
\caption{Execution Capabilities and Safety Comparison}
\label{tab:execution_safety_comparison}
\begin{tabular}{|l|p{2.2cm}|p{2.3cm}|}
\hline
\textbf{Property} & \textbf{Agentic Coding} & \textbf{Vibe Coding} \\
\hline
Code Execution & Autonomous, within containerized runtime environments & Manual execution via IDE or terminal \\
\hline
Testing Pipeline & Integrated test agents with recursive retry logic & Developer-initiated tests, often ad hoc \\
\hline
Feedback Loop & Embedded, goal-aware iteration with self-correction & Post-hoc, developer-controlled review and re-prompting \\
\hline
Security Enforcement & Namespace isolation, policy constraints, access controls & Relies on post-generation linting and external static analysis \\
\hline
Explainability & Full execution trace, rollback logs, action graph inspection & No native traceability; depends on developer recall or documentation \\
\hline
Prompt Sensitivity & Lower; robust to ambiguity through planning & High; vague prompts produce erratic or incomplete outputs \\
\hline
Ideal Use Cases & CI/CD automation, infra provisioning, large-scale refactoring & Rapid prototyping, exploratory design, educational coding \\
\hline
\end{tabular}
\end{table}

To illustrate these distinctions, consider the task of implementing a RESTful (\href{https://restfulapi.net/}{RESTful API} JWT-based authentication system. In a vibe coding workflow, the developer begins by prompting the LLM with a natural language instruction \cite{zamfirescu2025beyond, ma2025should}. The model generates a code snippet that directly reflects prior examples in its training data. For instance:

\begin{tcolorbox}[title=Generated Output: Vibe, colback=gray!5!white, colframe=gray!80!black]
\begin{verbatim}
from fastapi import Depends, 
HTTPException
from jose import JWTError, jwt

SECRET_KEY = "mysecret"

def verify_token(token: str):
    try:
        payload = jwt.decode(token, 
        SECRET_KEY)
        return payload
    except JWTError:
        raise HTTPException
        (status_code=403)
\end{verbatim}
\end{tcolorbox}

The developer must then validate this implementation manually using tools like Postman or `curl`, inspect error behavior, and refine the prompt if improvements are needed. This feedback loop remains external and human-driven.

In contrast, an agentic system initiates a structured execution pipeline \cite{zhang2025chainbuddy, ke2025survey}. The agent begins by analyzing route configurations, generates middleware for token validation, injects this into the FastAPI framework, executes automated tests via a test client, logs any exceptions, and autonomously retries modifications \cite{ali2025secure}. This closed-loop approach supports reproducibility and validation at every step.

\begin{tcolorbox}[title=Agentic Code Execution Flow, colback=blue!5!white, colframe=blue!75!black]
\begin{verbatim}
Step 1: Analyze API routes
Step 2: Generate token middleware
Step 3: Inject middleware 
        into FastAPI
Step 4: Run test client
Step 5: Log errors and fix 
        recursively
\end{verbatim}
\end{tcolorbox}

This illustration highlights the architectural divergence in action: vibe coding emphasizes developer-led exploration, whereas agentic coding emphasizes autonomous, testable construction. The former thrives in low-stakes or creative contexts; the latter aligns with enterprise-grade reliability and scalable automation. Vibe coding and agentic coding are not competing paradigms but represent complementary trajectories in AI-assisted software engineering. Understanding their technical architecture, capabilities, and constraints is essential for designing effective toolchains and selecting the appropriate paradigm based on context. The next section evaluates these models across performance, efficiency, and deployment scalability dimensions.

\section{Practical Workflow Differences}

The practical adoption of Vibe Coding and Agentic Coding paradigms reveals fundamental differences in developer interaction models, cognitive frameworks, workflow architectures, and application suitability. This section presents a comparative investigation across four dimensions: developer roles and mental models, workflow patterns, engagement modes, and human-system factors. Through illustrative examples and comparative tables, we outline how each paradigm supports different stages of software development, from rapid prototyping to automated refactoring and large-scale system integration.

\subsection{Developer Roles and Mental Models}

\subsubsection{Vibe Coding: Dialogic Creation and Exploratory Interaction}

Vibe Coding emphasizes an interactive, conversational dynamic between the developer and the LLM. Developers are engaged as co-creators, navigating design and implementation decisions through iterative prompt-response cycles \cite{yang2025code, hsu2025programming}. This approach lowers the activation threshold for idea exploration, enabling developers to articulate abstract requirements and progressively converge on working solutions.

\textbf{Primary Roles:}
\begin{itemize}
    \item \textbf{Intent Architect:} Formulates project goals in natural language, refining intent through prompt iteration.
    \item \textbf{Creative Director:} Evaluates, edits, and curates AI-generated outputs to align with design intent and user experience.
    \item \textbf{Explorer:} Uses the AI to experiment with unknown APIs, test UI patterns, or scaffold new features with minimal prior knowledge.
\end{itemize}

\textbf{Cognitive Model:} The developer operates with a “what-before-how” mindset articulating high-level needs (e.g., “Build a login page with 2FA”) and assessing the AI’s proposed structural and syntactic solutions. This model promotes rapid feedback and creative experimentation but delegates testing and validation responsibilities to the developer.

\begin{tcolorbox}[title=Example Prompt Loop, colback=white, colframe=black]
\textbf{Developer:} ``Create login API with password hashing and 2FA."

\textbf{AI:} Generates FastAPI endpoint using JWT + pyotp.

\textbf{Developer:} ``Add unit tests and refactor 2FA logic into middleware."
\end{tcolorbox}

\subsubsection{Agentic Coding: Task Delegation and Strategic Oversight}

Agentic Coding reframes the developer’s role as that of a systems architect, strategic planner, and supervisory reviewer. Developers define high-level tasks or objectives, which are parsed and decomposed by autonomous agents that execute software engineering workflows ranging from code modification to integration testing and version control \cite{jin2024llms, he2024llm}.

\textbf{Primary Roles:}
\begin{itemize}
    \item \textbf{Strategic Planner:} Specifies tasks, objectives, and architectural constraints for the agent to act upon \cite{acharya2025agentic, sapkota2025ai}.
    \item \textbf{Supervisor:} Monitors execution trace logs, performance reports, and system outputs \cite{moshkovich2025beyond}.
    \item \textbf{Reviewer:} Validates the correctness, maintainability, and security of agent-generated changes before integration \cite{luo2025large}.
\end{itemize}

\textbf{Cognitive Model:} Developers think in terms of orchestration rather than direct implementation. A single instruction such as “Fix broken login and ensure OAuth2 compliance” may be internally decomposed by the agent into authentication token migration, CI pipeline updates, test reruns, and dependency auditing. Human intervention is minimized to exception handling or ambiguity resolution.

\subsection{Workflow Patterns}

\subsubsection*{Vibe Coding: Conversational Exploration}

Vibe coding workflows are inherently exploratory and non-linear. Developers issue prompts, inspect generated code, and provide incremental feedback \cite{nam2025prompting, cui2025tests}. This model is optimal for interface prototyping, low-risk experimentation, or knowledge discovery \cite{wu2025ai}.

\textbf{Example – Dashboard Prototyping}
\begin{enumerate}
    \item Developer: ``Build React dashboard with user count, revenue, and churn chart."
    \item AI: Generates UI with Chart.js and dummy data.
    \item Developer: ``Add tooltips and export to CSV."
    \item AI: Adds hover logic and export buttons.
    \item Developer: ``Write Cypress tests."
    \item AI: Outputs E2E test coverage.
\end{enumerate}

\subsubsection*{Agentic Coding: Structured Execution Pipelines}

Agentic coding follows structured workflows based on task planning \cite{niu2025flow, caetano2025agentic}, state management \cite{ding2025frontend}, and recursive feedback loops \cite{dawid2025agentic, buehler2025preflexor}. These workflows suit enterprise-grade tasks requiring correctness, traceability, and automation.

\textbf{Example – Automated Dependency Upgrade}
\begin{enumerate}
    \item Developer: ``Upgrade all npm packages to latest secure versions."
    \item Agent:
    \begin{itemize}
        \item Parses \texttt{package.json}
        \item Updates dependency versions
        \item Executes test suite
        \item Resolves compatibility issues
        \item Generates changelog
    \end{itemize}
    \item Developer: Reviews logs and approves pull request.
\end{enumerate}

\subsection{Comparative Analysis: Developer Engagement and Workflow Suitability}
The differing interaction paradigms of Vibe Coding and Agentic Coding are reflected not only in architectural and cognitive models, but also in their practical workflow characteristics. From the role of the developer and interaction patterns to testing, documentation, and error resolution, each paradigm supports distinct modes of software creation. These differences have significant implications for project scale, team composition, and toolchain integration. Table~\ref{tab:workflowcompare} presents a structured comparison of key workflow dimensions, highlighting how each model aligns with specific development contexts and use cases, and guiding practitioners in selecting the appropriate strategy for their software engineering objectives.

\begin{table}[htbp]
\centering
\caption{Developer Roles and Workflow Comparisons}
\label{tab:workflowcompare}
\begin{tabular}{|l|p{2.2cm}|p{2.5cm}|}
\hline
\textbf{Dimension} & \textbf{Vibe Coding} & \textbf{Agentic Coding} \\
\hline
Developer Role & Intent Architect, Creative Partner & Supervisor, Strategic Planner \\
\hline
Workflow Type & Conversational, Fluid & Structured, Recursive \\
\hline
Interaction Style & Prompt-response loop & Task delegation and logging \\
\hline
Best Use Case & Rapid Prototyping, UX Design & Refactoring, Maintenance, QA \\
\hline
Learning Curve & Low (natural language) & Moderate (agent supervision) \\
\hline
Testing & Manual or prompt-based & Automated and integrated \\
\hline
Documentation & On-demand via prompt & Auto-generated during execution \\
\hline
Error Resolution & Developer-guided & Agent-driven with fallback logic \\
\hline
\end{tabular}
\end{table}

\subsection{Scientific and Human Factors}

\subsubsection*{Cognitive Load and Developer Productivity}

\textbf{Vibe Coding} reduces cognitive load associated with syntax and implementation details \cite{kazemitabaar2023novices, fakhoury2024llm}, enabling rapid ideation and creative flow. It is especially effective for solo developers, early prototyping, or teaching new frameworks through interaction.

\textbf{Agentic Coding} introduces new cognitive demands in terms of system understanding, trust calibration, and supervision. However, it scales well in complex systems, enabling experienced developers to manage multiple asynchronous workflows and integrate formal validation into the pipeline \cite{bobko2023human, secchi2024complexity, tariverdi2024trust}.

\subsubsection*{Collaboration and Team Models}

\textbf{Vibe Coding} is well-suited to collaborative scenarios such as hackathons or pair programming. Multiple developers may interact with the same agent in conversational loops to co-create ideas.

\textbf{Agentic Coding} enables distributed responsibility across modular systems. Individual agents or agent groups may be assigned to subsystem-level tasks, supporting parallelism and pipeline scalability in team-based development.

\subsection{Real-World Scenarios}

\subsubsection*{Vibe Coding – Social Media Feature Ideation}

To illustrate the fluid, creative, and iterative nature of vibe coding, consider the development of a new feature in a mobile social media application. In this workflow, the developer incrementally guides the LLM through a series of conversational prompts, shaping both frontend and backend functionality while maintaining control over design decisions. The example below demonstrates how a story highlight feature is ideated, implemented, refined, and tested using natural language interaction.

\begin{tcolorbox}[title=Vibe in Practice: Social Highlights, colback=green!5!white, colframe=green!75!black]
\textbf{Prompt:} ``Add story highlight feature to mobile app."

\textbf{AI:} Creates backend schema, API endpoints, frontend UI (React Native).

\textbf{Follow-up:} ``Allow drag-and-drop ordering."

\textbf{AI:} Adds sortable list with gesture control.

\textbf{Prompt:} ``Write tests."

\textbf{AI:} Outputs Jest + E2E test scripts.
\end{tcolorbox}

\subsubsection*{Agentic Coding – Legacy Migration}

In contrast, agentic coding systems are designed for high-autonomy, structured workflows. The following example showcases an agentic system tasked with migrating a legacy codebase from Python 2 to Python 3. Once the developer issues the high-level instruction, the agent autonomously scans the codebase, applies systematic refactoring, validates changes through testing, and reports results for human approval. This highlights how agentic systems can automate extensive codebase transformations with minimal oversight.

\begin{tcolorbox}[title=Agentic in Practice: Python 2 to 3 Migration, colback=green!5!white, colframe=green!75!black]
\textbf{Prompt:} ``Upgrade all Python code to version 3.x."

\textbf{Agent:}
\begin{itemize}
    \item Parses source files for deprecated syntax.
    \item Applies automatic and rule-based refactoring.
    \item Executes test suite.
    \item Reports diffs and unresolved issues.
\end{itemize}

\textbf{Developer:} Reviews logs, patches edge cases, and approves merge.
\end{tcolorbox}

Vibe Coding and Agentic Coding represent two ends of the AI-assisted software development spectrum. Vibe Coding excels in human-in-the-loop, creative workflows where rapid feedback and flexibility are essential \cite{iyenghar2025clever,haque2025llms, sabbah2025humans}. Agentic Coding, by contrast, emphasizes autonomy, reliability, and integration making it ideal for large-scale automation and enterprise deployment \cite{acharya2025agentic, sapkota2025ai}. Understanding the operational strengths and trade-offs of both paradigms and hybridizing them effectively offers a pathway toward more intelligent, efficient, and adaptive software development ecosystems.

\section{Implementation Strategies}
The implementation strategies of Vibe Coding and Agentic Coding diverge fundamentally in how they translate developer instructions into executable software. While both rely on LLMs, the strategies they adopt for prompt handling, code verification, and tool integration reflect distinct philosophies. This section explores these dimensions in detail.

\subsection{Prompt Engineering}

\subsubsection*{Vibe Coding: Precision and Context for Creative Generation}
Prompt engineering is the backbone of effective vibe coding. Here, developers provide explicit, intent-focused natural language prompts to guide the AI model in generating desired code outputs \cite{gu2023systematic, geroimenko2025essential}. The process is inherently iterative and dialogic \cite{robino2025conversation, velasquez2023prompt}.

\textbf{Key Characteristics:}
\begin{itemize}
    \item \textbf{Intent-Focused:} Prompts emphasize desired outcomes rather than procedures \cite{schulhoff2024prompt, robertson2024game}.
    \item \textbf{Context-Sensitive:} LLMs use extended context windows (16k--32k tokens) to retain semantic awareness of the codebase \cite{xu2025clover, chen2024long, zhang2024chain}.
    \item \textbf{Exploratory:} Prompts may include multiple solution requests or alternative suggestions \cite{mondal2024enhancing, cao2025study}.
\end{itemize}

\begin{tcolorbox}[title=Example: Iterative Prompt Refinement, colback=orange!5!white, colframe=orange!75!black]
\textbf{Developer:} "Create a REST API with user login and JWT auth."

\textbf{AI:} Returns FastAPI endpoint with pyjwt integration.

\textbf{Developer:} "Add 2FA with OTP."

\textbf{AI:} Adds OTP-based validation logic with PyOTP.
\end{tcolorbox}

Vague prompts such as "Make this more secure" often result in ambiguous or generic outputs. Specific prompts significantly improve precision.

\subsubsection*{Agentic Coding: Hierarchical and Multi-Step Instructions}

Agentic coding requires hierarchical prompting suited to multi-phase task execution \cite{fourney2024magentic, buehler2025preflexor, sapkota2025ai}. Developers issue macro-level instructions that agents deconstruct into subtasks \cite{kant2025develop}.

\textbf{Key Characteristics:}
\begin{itemize}
    \item \textbf{Task Chaining:} Prompts are designed to be decomposable.
    \item \textbf{Execution-Ready:} Prompts include constraints, file references, and expected outputs.
    \item \textbf{Interactive Feedback:} Agents may prompt for clarification before execution.
\end{itemize}

\begin{tcolorbox}[title=Example: Agentic Prompt Workflow, colback=orange!5!white, colframe=orange!75!black]
\textbf{Prompt:} ``Update all modules using Flask 1.x to Flask 2.x and resolve deprecation warnings."

\textbf{Agent Actions:}
\begin{enumerate}
    \item Parse imports and deprecated patterns.
    \item Replace and test each submodule.
    \item Run CI pipeline.
    \item Log any exceptions or failed tests.
    \item Compile a migration report.
\end{enumerate}
\end{tcolorbox}

Effective agentic prompts must encode enough structure to support independent agent decision-making while retaining flexibility for adaptive correction.

\subsection{Review and Debugging}

Debugging and validation workflows are critical in shaping the reliability and usability of AI-assisted software development. Vibe Coding and Agentic Coding differ fundamentally in how error detection, correction, and verification are handled. Vibe Coding emphasizes human-in-the-loop inspection, relying on manual review and iterative prompt refinement, making it conducive to exploratory development. In contrast, Agentic Coding shifts much of this responsibility to autonomous agents equipped with runtime monitoring, log analysis, and rollback mechanisms. This section systematically analyzes the validation strategies, developer effort, and debugging affordances inherent in each paradigm.

\subsubsection{Vibe Coding: Manual, Iterative, and Prompt-Based Validation}

In Vibe Coding, the developer plays an active role in identifying and resolving issues. Errors are detected through manual testing, visual inspection, and ad hoc interaction with the LLM. The absence of a built-in feedback mechanism means the AI does not autonomously detect or act upon code failures. Instead, the developer diagnoses errors and re-engages the model with revised prompts.

\textbf{Key Techniques:}
\begin{itemize}
    \item \textbf{Post-Generation Review:} Developers visually examine AI-generated code, sometimes assisted by static analysis tools or linters.
    \item \textbf{Manual Testing:} Code is copied into a local IDE or REPL, where tests are written and executed manually.
    \item \textbf{Prompt Debugging:} Errors are addressed by refining the original prompt or issuing follow-up questions to isolate root causes.
\end{itemize}

\begin{tcolorbox}[title=Example: Prompt-Based Debugging, colback=green!5!white, colframe=green!70!black]
\textbf{Developer:} ``Why does this throw a 403 error?"

\textbf{AI:} ``JWT token decode fails due to missing audience field. Add 'aud' claim."
\end{tcolorbox}

While time-consuming, this approach fosters a high degree of developer engagement and intuition building. It is particularly effective in early-stage prototyping, learning new libraries, or customizing third-party tools, where interpretability and control are prioritized over automation.

\subsubsection{Agentic Coding: Autonomous Debugging and Runtime Verification}

Agentic Coding systems are equipped with autonomous verification mechanisms embedded into their runtime architecture \cite{dennis2016practical, luckcuck2019formal, dennis2020verifiable}. Once a task is defined, the agent is capable of not only executing it but also validating its correctness through runtime evaluation and internal logging systems (using systems such as runtime verifications \cite{havelund2015rule, francalanza2018runtime, jennings2003agent}. These systems often include error-handling protocols such as rollback, patch substitution, or retry logic.

\textbf{Key Techniques:}
\begin{itemize}
    \item \textbf{Runtime Evaluation:} Tasks are executed in secure sandboxes or containers simulating production environments.
    \item \textbf{Log Inspection:} Execution traces, diffs, and system states are recorded and used for post-mortem analysis or decision trees.
    \item \textbf{Error Rollback:} On failure, agents automatically revert to a prior stable state or reattempt execution with modified inputs.
\end{itemize}

\begin{tcolorbox}[title=Agentic Debug Loop, colback=blue!5!white, colframe=blue!75!black]
\textbf{Agent Workflow:}
\begin{itemize}
    \item Refactor class structure.
    \item Run full test suite.
    \item On failure: isolate diff, revert, and retry with alternate patch.
    \item Finalize changes only if all tests pass.
\end{itemize}
\end{tcolorbox}

This feedback-driven model supports CI/CD pipelines, automated patching, and enterprise-grade refactoring. It minimizes manual intervention, ensures reproducibility, and scales efficiently across larger codebases. However, it requires careful oversight to validate that the agent’s autonomous decisions align with project constraints and security standards.

\subsection{Tool Ecosystems}

The tooling landscape supporting Vibe Coding and Agentic Coding reflects the fundamental differences in their architectural goals and user interaction models. Vibe Coding tools are designed to support rapid prototyping, fluid human-AI interaction, and minimal onboarding, often operating within lightweight development environments. In contrast, Agentic Coding platforms are engineered to manage autonomous task execution, system orchestration, and compliance logging, often integrating deeply with infrastructure-level operations. This section summarizes key tools representative of each paradigm and their associated capabilities.

\subsubsection*{Vibe Coding Tools: Conversational Interfaces and Creative Assistants}

Vibe coding tools prioritize accessibility, immediacy, and seamless integration into the developer’s creative flow. These platforms typically use prompt-response loops and are embedded within IDEs, browsers, or chat interfaces, enabling on-the-fly code generation, inline explanations, and lightweight context sharing.

\textbf{Representative Platforms:}
\begin{itemize}
    \item \textbf{ChatGPT:} Provides flexible, conversational code generation and prompt debugging via natural language input \cite{achiam2023gpt}.
    \item \textbf{Gemini:} Combines code suggestion capabilities with contextual documentation references and visual UI integration \cite{team2023gemini}.
    \item \textbf{Claude (Anthropic):} Offers long-context code assistance with a focus on safe interaction and persistent multimodal reasoning \href{https://www.anthropic.com/claude}{Claude anthropic}.
    \item \textbf{Replit Ghostwriter:} Embeds a code-generating LLM into a browser-native IDE for real-time, context-aware assistance \href{https://replit.com/learn/intro-to-ghostwriter}{Replit}.
\end{itemize}

These tools excel in tasks requiring quick iteration, explanation, and ideation, supporting solo developers, learners, and early-phase design exploration.

\subsubsection*{Agentic Coding Tools: Execution-Aware Agents and Autonomous Pipelines}

Agentic coding tools embed intelligent agents within controlled execution environments, enabling autonomous code transformation, validation, and deployment. These platforms often integrate with containers, logs, and permission systems to support continuous development and robust system-level interaction.

\textbf{Representative Platforms:}
\begin{itemize}
    \item \textbf{Codex (OpenAI):} Supports full project-level task execution, including file system manipulation, test orchestration, and CI pipeline triggering in sandboxed cloud instances \href{https://openai.com/codex/}{Codex OpenAI Link}.
    \item \textbf{Claude Code (Anthropic):} Designed with explainability and oversight features, enabling developers to audit changes, trace reasoning, and rollback unsafe actions \href{https://www.anthropic.com/claude-code}{ Claude Code Link}.
    \item \textbf{Jules:} Google Jules is an autonomous AI coding agent powered by Gemini that handles tasks like bug fixes and feature updates asynchronously \href{https://jules.google/}{Google Jules}. It operates on cloned GitHub repos in secure VMs, presenting results for review before integration.
\end{itemize}

These agentic tools are suited for scenarios demanding auditability, security enforcement, and task autonomy across large and distributed codebases.

\textbf{Comparative Analysis: Implementation Strategies}
The practical implementation strategies of Vibe Coding and Agentic Coding reflect their divergent philosophies in autonomy, tooling, and workflow integration. While Vibe Coding emphasizes rapid ideation, creative prompting, and developer-centered iteration, Agentic Coding systems are designed for structured task decomposition, automated execution, and autonomous debugging. These contrasts are not merely stylistic but architectural, influencing everything from execution environments and review cycles to tool selection and deployment suitability. Table~\ref{tab:impl_strat_compare} provides a comparative overview of the key implementation dimensions, offering a concise synthesis of how each paradigm operationalizes AI-assisted software development across prototyping, debugging, and production-scale maintenance tasks.

\begin{table}[htbp]
\centering
\caption{Implementation Strategy Comparison: Vibe vs. Agentic}
\begin{tabular}{|l|p{2.2cm}|p{2.3cm}|}
\hline
\textbf{Aspect} & \textbf{Vibe Coding} & \textbf{Agentic Coding} \\
\hline
Prompt Engineering & Intent-based, iterative, creative & Hierarchical, task-oriented, detailed \\
\hline
Debugging & Manual, ad hoc, prompt-driven & Automated, agent-controlled \\
\hline
Execution Context & External IDEs, local testing & Sandbox, container runtime \\
\hline
Tool Examples & ChatGPT, Gemini, Replit & Codex, Claude Code, Roo \\
\hline
Review Cycle & Developer-inspected & Agent-audited + logs \\
\hline
Suitability & Prototyping, experimentation & Maintenance, integration, enterprise-scale ops \\
\hline
\end{tabular}
\label{tab:impl_strat_compare}
\end{table}

\subsection{Scientific and Practical Implications}

The application of Vibe Coding and Agentic Coding reflects two distinct paradigms in AI-assisted software development. Vibe Coding facilitates rapid ideation and creative exploration through lightweight, prompt-driven workflows. Its primary strengths lie in minimal setup time, intuitive experimentation, and support for learning new tools or frameworks. However, its limitations such as the need for manual validation, reduced suitability for production environments, and lack of built-in quality assurance make it less ideal for scalable engineering tasks.

In contrast, Agentic Coding emphasizes automation, structure, and reliability. Agents can autonomously perform task planning, testing, and code integration, reducing developer burden and increasing QA compliance. These systems are well-suited for CI/CD pipelines, legacy modernization, and enterprise-level code maintenance. Still, agentic workflows require secure execution environments, complex orchestration infrastructure, and careful oversight to prevent silent failures from misconfigurations.

Ultimately, these paradigms are not mutually exclusive. Vibe Coding is optimal during early-stage design and prototyping, while Agentic Coding excels in implementation and operational phases. Hybrid workflows that leverage both beginning with creative prompting and transitioning to autonomous execution can maximize efficiency and robustness. As AI coding agents evolve, this blended strategy will likely define the next generation of intelligent development practices.

\section{Use Cases and Applications}

The evolution of AI-assisted software development has led to the emergence of two complementary paradigms: Vibe Coding and Agentic Coding. Each presents distinct strengths, ideal workflow conditions, and technical affordances. Understanding where each paradigm excels enables developers and organizations to align tooling strategies with project goals, lifecycle stages, and team expertise. This section explores the optimal application scenarios for both approaches based on functional characteristics, observed benefits, and empirical use. A comparative summary of use case suitability is provided in Table~\ref{tab:usecase_compare}.

\subsection{Ideal Scenarios for Vibe Coding}

Vibe Coding is most effective in early-stage development and exploratory contexts where creativity, rapid feedback, and flexible control are essential. It supports three dominant use cases:

\textit{Creative Exploration:} Vibe tools allow developers to experiment with high-level design ideas, generate multiple implementation variants, and visualize UI flows or logic strategies in real time. This makes them ideal for brainstorming sessions, algorithmic sandboxing, and feature sketching.

\textit{Rapid Prototyping:} Vibe coding excels in building functional MVPs from natural language prompts. Developers can generate full-stack components including REST APIs, frontend layouts, and test cases within hours using conversational iterations, making it suitable for agile environments and hackathons.

\textit{Learning New Technologies:} For onboarding or upskilling, vibe tools serve as real-time coding tutors. Developers can query LLMs for explanations, receive context-sensitive code snippets, and troubleshoot unfamiliar tools or frameworks like Next.js, WebAssembly, or Rust interactively.

\subsection{Ideal Scenarios for Agentic Coding}

Agentic Coding is engineered for structured, high-reliability workflows where automation and scale are critical. It is best applied in production environments and enterprise operations:

\textit{Codebase Refactoring:} Agentic tools can autonomously analyze legacy systems, detect outdated code patterns, and apply systematic refactorings. For example, migrating a Python 2.7 system to Python 3.x involves syntax updates, dependency rewrites, and full test coverage all achievable with minimal human oversight using agentic pipelines.

\textit{Routine Engineering Tasks:} These include automated dependency upgrades, code formatting, test regeneration, and CI/CD pipeline maintenance. Agents apply consistent standards across large codebases, improving maintainability and reducing manual engineering overhead.

\textit{Regression Bug Fixing:} Agentic systems excel at log-based error diagnosis, root-cause analysis, and autonomous code repair. They reduce mean time to resolution (MTTR) by running test suites, applying patches, and updating changelogs without developer intervention ideal for mission-critical services.

\subsection{Comparative Analysis: Vibe vs. Agentic in Practice}

\begin{table}[htbp]
\centering
\caption{Practical Use Case Comparison of Agentic vs. Vibe Coding}
\label{tab:usecase_compare}
\begin{tabular}{|l|p{2.5cm}|p{2.6cm}|}
\hline
\textbf{Scenario} & \textbf{Agentic Strengths} & \textbf{Vibe Strengths} \\
\hline
Codebase Refactoring & Autonomous, scalable, test-integrated & Limited to small, manual edits \\
\hline
Routine Tasks & Fully automated and uniform & Good for trial runs and local edits \\
\hline
Regression Fixes & Log-driven triage, self-healing agents & Useful for hands-on debugging practice \\
\hline
Creative Exploration & Predictable, less adaptive & Highly responsive, freeform ideation \\
\hline
Prototyping & Backend-ready automation & End-to-end generation from prompts \\
\hline
Learning & Fast setup, context recall & Interactive Q\&A, stepwise clarification \\
\hline
\end{tabular}
\end{table}

\subsection{Real-World Applications}

\subsubsection*{Vibe Coding: 10 Practical Use Cases}
\begin{figure*}
    \centering
    \includegraphics[width=0.85\linewidth]{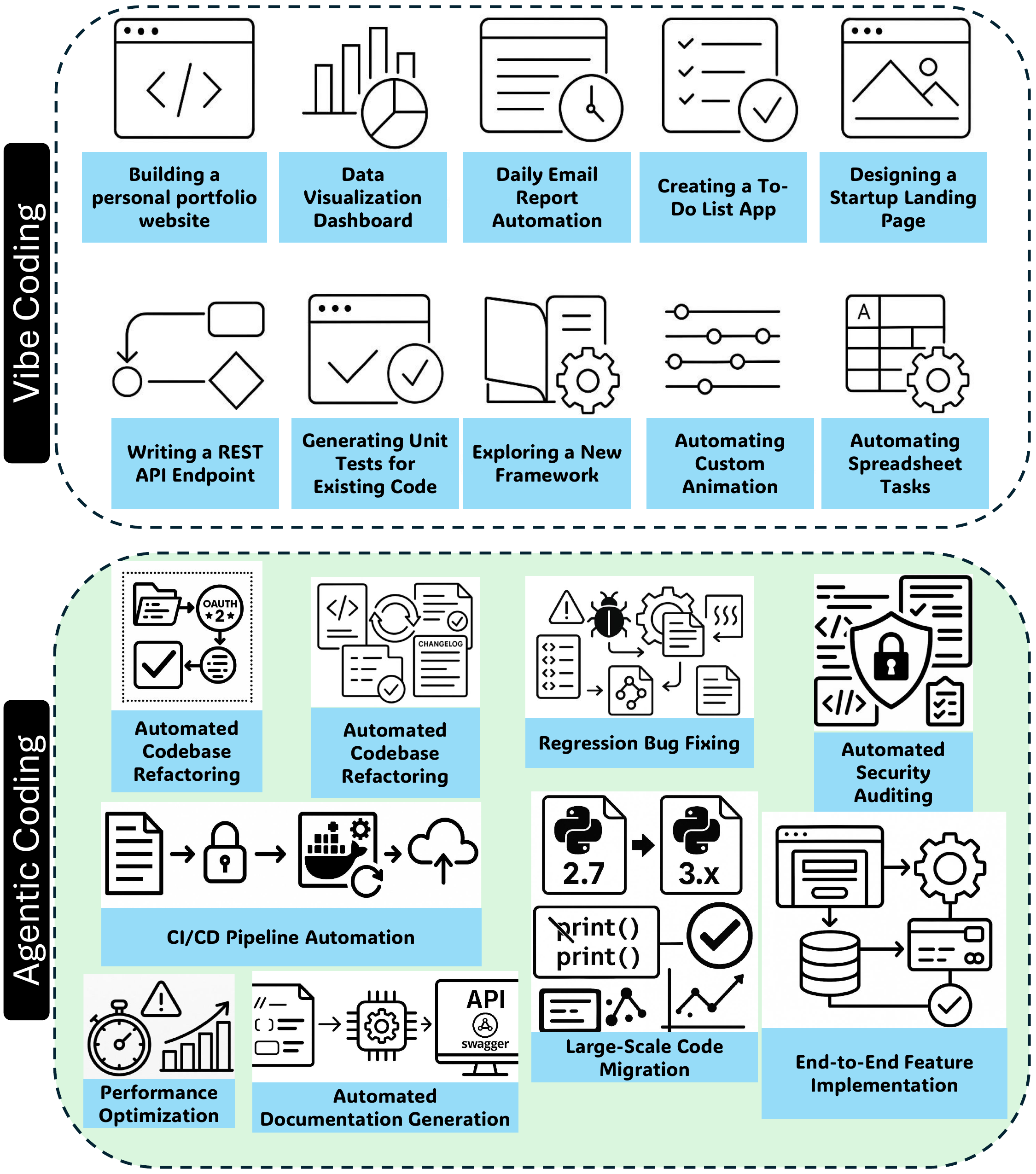}
    \caption{10 use cases of Vibe Coding (Upper part) and Agentic Coding (Lower Part)}
    \label{fig:examples}
\end{figure*}

Figure \ref{fig:examples} illustrated the summary of 10 practical use cases of vibe coding, and each use cases are detailed into following points:
\begin{enumerate}

\item \textbf{Personal Portfolio Website Development}

Vibe coding proves highly effective in generating professional personal websites with minimal manual effort \cite{yuan2025ui2html, toth2024llms, xu2025web}. For instance, a developer may prompt, “Create a modern, responsive personal website with sections for About, Projects, and Contact. Use React and include a dark mode toggle.” The AI interprets this instruction and outputs a full React-based project including reusable components, routing with React Router, state management for theme toggling, and styled-components for UI design. Importantly, the AI-generated code adheres to modern frontend architecture patterns, enabling responsiveness across screen sizes and semantic markup for accessibility. Developers can then refine colors, branding, or layout logic without spending time on boilerplate setup. This use case exemplifies how vibe coding empowers solo developers, students, or freelancers to launch polished portfolios in hours, promoting self-representation, employment outreach, or client engagement. Furthermore, it enables real-time customization: prompts like “Add a testimonials section with a carousel” can be appended iteratively. The ease of interaction and semantic clarity of natural language prompts transform what would traditionally take several days of setup, CSS tweaking, and component reuse into a streamlined, conversational workflow making this a flagship use case in frontend ideation through Vibe Coding \cite{esposito2024programming, johnsen2025developing}.

\item \textbf{Interactive Data Visualization Dashboards}

Another powerful use case for vibe coding lies in developing interactive data dashboards \cite{ xu2025powerpoint}. A prompt such as “Build an interactive dashboard that displays sales data as a bar chart and a pie chart, with filters for region and date” activates the model’s ability to generate full JavaScript-based UI with integrated visualization libraries like Chart.js or D3.js. The AI constructs components to handle input state (e.g., dropdowns or sliders for region and date), binds them to data filters, and connects the outputs to responsive visual elements. This approach significantly accelerates prototyping for data scientists, product managers, or researchers who may lack the frontend expertise to translate insights into live web interfaces. Moreover, vibe coding allows iterative updates such as “Add a line chart for monthly revenue trends” or “Include export to CSV functionality,” enabling flexible expansion without reworking foundational architecture. These dashboards can be powered by static JSON or API-backed data sources, depending on the context. The AI-generated layout often includes accessibility features, tooltips, and mobile responsiveness by default. This use case demonstrates the integration of data fluency and visual storytelling \cite{shen2024data, shao2025narrative}, helping domain experts bridge the gap between backend analytics and user-facing visualizations using conversational coding workflows.

\item \textbf{Daily Email Report Automation}

Vibe coding excels in automating routine workflows such as scheduled email reports \cite{wang2024executable, joshi2024coding}. Given a prompt like “Write a Python script that pulls yesterday’s sales from a CSV file and emails a summary to my team at 8am every day,” the AI generates code using Python libraries including \texttt{pandas} for CSV parsing, \texttt{smtplib} for sending emails, and built-in modules like \texttt{datetime} for filtering data. The script typically formats metrics into human-readable summaries, attaches files if necessary, and configures recipient lists and message subjects. Beyond the core logic, the AI often includes instructions for setting up a cron job or a Windows Task Scheduler entry to automate execution. This workflow is particularly useful for small businesses, sales teams, or researchers who need daily reporting but lack access to enterprise-grade BI platforms. Further enhancements like “Format summary as HTML table” or “Attach CSV of filtered data” can be integrated seamlessly through subsequent prompts. Compared to traditional scripting workflows that require detailed knowledge of syntax and library APIs, vibe coding reduces setup time and encourages iterative improvements \cite{taulli2024ai, zhang2024large}. This use case exemplifies how AI-enhanced scripting democratizes automation, enabling non-expert programmers to implement robust reporting pipelines through natural language commands.

\item \textbf{To-Do List Web Application}

Vibe coding provides a streamlined approach to developing interactive, stateful web applications, such as to-do list managers \cite{voronin2024development, trivedi2024appworld}. Given the prompt “Make a simple to-do list web app with add, remove, and mark-as-complete features. Use Vue.js,” the AI generates a project that includes reusable Vue components for task entry, task listing, and status toggling. It sets up reactive data binding via the Composition or Options API and persists the task list using browser localStorage or sessionStorage. This use case is particularly effective for frontend learners, rapid prototyping, and UI logic testing, where state management and component communication are essential. Developers can incrementally add features like task categorization, due dates, or UI transitions simply by prompting further e.g., “Add color-coded categories for each task” or “Sort tasks by due date.” Importantly, the AI typically follows modern Vue conventions, using scoped CSS, modular scripts, and semantic HTML structure. This hands-on feedback loop fosters both conceptual understanding and production ready outcomes \cite{rasheed2024codepori}. Furthermore, the generated application can be easily deployed using platforms like Vercel \href{https://vercel.com/}{Vercel Link} or Netlify \href{https://www.netlify.com/}{Netlify Link}, showcasing vibe coding’s low barrier to full-cycle application development. In short, to-do list apps serve as an ideal sandbox to observe how AI interprets dynamic user interfaces and local state orchestration from natural language descriptions.

\item \textbf{Startup Landing Page Generation}

Vibe coding significantly accelerates the creation of marketing-oriented landing pages an essential component for startups \cite{dale2024start}, product demos \cite{chew2023llm, tabarsi2025llms}, and digital campaigns \cite{sniegowski2025evaluating}. A prompt such as “Generate a landing page for a new AI-powered note-taking app. Include a hero section, features, testimonials, and a signup form” guides the AI to output semantically structured HTML and Tailwind CSS with clearly delineated sections. It generates layouts with responsive flexbox or grid-based positioning, embedded SVG icons, call-to-action (CTA) buttons, and form validation logic. The landing page is typically annotated with placeholder text and sample imagery, which the developer can replace for customization. This use case benefits non-technical founders or small teams seeking to deploy marketing content quickly without hiring web designers. Moreover, it allows iterative refinement: prompts like “Add newsletter opt-in with Mailchimp integration” or “Include pricing tiers with toggleable monthly/annual plans” are easily interpreted by the model. The AI’s outputs align with SEO best practices and accessibility guidelines, often including meta tags and ARIA attributes. This workflow illustrates how vibe coding transforms vague branding concepts into fully deployable pages, enabling rapid iteration on visual identity and user acquisition strategies through natural language co-design.

\item \textbf{RESTful API Endpoint Development}

Vibe coding is increasingly effective for backend prototyping, particularly in generating modular RESTful APIs. A prompt like “Create a Node.js Express endpoint for user registration, with email validation and password hashing” leads to the generation of structured middleware logic using \texttt{express}, \texttt{bcryptjs}, and \texttt{validator.js}. The AI often scaffolds the file structure, initializes an \texttt{app.js} or \texttt{index.js}, and integrates JSON body parsing with standardized error handling. The resulting endpoint typically includes robust input validation, secure password storage, and informative response messages. This use case is particularly useful for full-stack developers prototyping backend logic without investing in boilerplate or configuration overhead. The generated code is modular enough to extend with authentication middleware like JWT, rate limiting, or database integration through MongoDB or PostgreSQL. Subsequent prompts e.g., “Add input sanitization to prevent XSS” or “Integrate this with a MongoDB user model” can be layered seamlessly. This shows how vibe coding supports iterative back-end design with conversational refinement. More importantly, it empowers frontend-focused developers to extend their capabilities into API development, promoting end-to-end skill integration. Thus, REST endpoint generation exemplifies how vibe coding reduces friction in backend service scaffolding, enhancing both speed and security of early-stage API development.

\item \textbf{Unit Test Generation for Frontend Components}

Testing is often overlooked in early-stage development, but vibe coding offers a frictionless approach to generating unit test suites for React and other component-driven frameworks. A prompt like “Write Jest unit tests for this React component that displays user profiles” initiates code that tests lifecycle methods, conditional rendering, prop validation, and event handling. The AI typically produces test files using \texttt{react-testing-library} or \texttt{enzyme}, defining mocks and asserting DOM state after simulated interactions. This functionality is invaluable for improving code coverage, debugging regressions, and maintaining confidence during refactors. Developers can extend the coverage by prompting “Add tests for error boundaries” or “Include mock API calls.” The generated test cases are often aligned with industry best practices, using descriptive test names and clear assertions. This use case demonstrates how vibe coding promotes quality assurance practices even in fast-paced prototyping contexts. It is particularly valuable for teams adopting test-driven development (TDD) or onboarding new engineers to legacy systems who need to create regression tests. By automating repetitive scaffolding of test logic, vibe coding helps enforce consistent testing structures, reducing the manual effort typically required for frontend validation and increasing reliability in component-driven architectures.

\item \textbf{Framework Exploration and Onboarding}

Vibe coding serves as an effective tool for developers exploring unfamiliar frameworks or ecosystems. For instance, a prompt such as “Show me how to set up a basic blog with Next.js, including routing and markdown support” yields a full scaffold of a modern web project. The AI typically generates key files including \texttt{pages/index.js}, dynamic routing components using \texttt{getStaticPaths} and \texttt{getStaticProps}, and integrates Markdown rendering through libraries like \texttt{remark} or \texttt{gray-matter}. It also provides file structures, package dependencies, and minimal working examples for content loading and layout templating. This guided setup facilitates hands-on learning without requiring prolonged documentation review, making it particularly beneficial for bootcamp students, self-taught developers, or engineers transitioning to unfamiliar stacks. Follow-up prompts such as “Add syntax highlighting for code blocks” or “Implement tag-based post filtering” can be layered incrementally, simulating real-world development flow. This workflow encourages immediate experimentation, rapid iteration, and experiential learning. Unlike tutorials or static documentation, vibe coding fosters an interactive feedback loop that enhances conceptual understanding through example-driven guidance. Thus, framework exploration via vibe coding accelerates onboarding while empowering users to transition from novice to productive contributor in modern development ecosystems like Next.js, SvelteKit, or Astro.

\item \textbf{Interactive Multimedia and Animation Prototyping}

Another high-value application of vibe coding lies in creating rich, interactive multimedia experiences. Given a prompt such as “Build a JavaScript animation that reacts to music and user clicks, with smooth transitions and colorful visuals,” the AI constructs a canvas-based or WebGL animation pipeline using libraries such as \texttt{p5.js}, \texttt{Tone.js}, or raw \texttt{requestAnimationFrame} logic. It wires real-time audio input events to visual transformations and handles user interactions like mouse movement or click-based effects. The output typically includes smoothing functions, frame buffers, and conditional rendering for responsiveness. This use case is highly relevant for frontend engineers, game developers, and digital artists looking to prototype immersive interfaces or creative installations. Because such applications are rarely template-driven, vibe coding’s strength lies in enabling fast ideation loops developers can iterate by prompting “Make colors change with beat intensity” or “Add particle trails to click animations.” This approach significantly lowers the barrier to generative art, sound visualization, or interactive infographics, domains where traditionally high skill thresholds once constrained creative experimentation. Vibe coding thus democratizes access to dynamic front-end graphics programming, enabling more developers to explore creative coding with immediacy and expressive control through natural language.

\item \textbf{Spreadsheet Automation with Google Apps Script}

Vibe coding extends beyond frontend and API development into automating productivity tools, such as spreadsheets. A use case like “Write a Google Apps Script to automatically color rows in a Google Sheet based on the value in the ‘Status’ column” triggers generation of JavaScript code tailored for the Apps Script environment. The model produces event-driven scripts using \texttt{onEdit(e)} handlers that evaluate cell values and apply conditional formatting via \texttt{setBackground()} methods. It also includes logic to optimize performance (e.g., range limiting) and optional enhancements like logging or undo triggers. This application is especially useful for educators, analysts, and administrative staff who manage large datasets and require visual cues for prioritization e.g., coloring tasks as “Complete,” “In Progress,” or “Overdue.” Additional prompts such as “Send email when status is ‘Blocked’” or “Sort by last updated date on edit” can expand the automation scope. Unlike manual scripting which requires prior knowledge of Google Apps Script APIs, vibe coding allows users to build task-specific automation routines through intuitive prompts. This use case illustrates how conversational AI can streamline routine digital workflows in office suites, bridging the gap between traditional spreadsheet usage and low-code enterprise automation.

\end{enumerate}

\subsubsection*{Agentic Coding: 10 Applied Use Cases}

\begin{enumerate}

\item \textbf{Automated Codebase Refactoring}

Agentic coding excels at large-scale, systematic code transformation, especially in legacy modernization scenarios. For example, given the instruction “Refactor all legacy authentication code to use OAuth2, update related tests, and ensure backward compatibility,” the agent parses the authentication module across files, identifies deprecated authentication logic, and systematically replaces it with OAuth2-compliant handlers. It updates environment configurations, rewrites affected middleware, and adjusts API headers to match new security flows. Unit and integration tests are updated or regenerated to validate the changes, and backward compatibility layers are introduced where applicable. Finally, the agent commits these changes to a Git branch and submits a changelog for review. This minimizes manual intervention and supports safer refactorings in production-sensitive codebases, where human error could be costly. Agentic refactoring is especially suitable for frameworks undergoing deprecation, major version upgrades, or compliance updates (e.g., from cookie-based sessions to token-based authentication).

\item \textbf{Routine Dependency Updates}

Maintaining up-to-date dependencies across large repositories is tedious and error-prone an ideal task for agentic automation. When prompted with “Update all project dependencies to their latest secure versions, fix any compatibility issues, and document changes,” the agent examines \texttt{package.json}, \texttt{requirements.txt}, or equivalent manifest files and upgrades each package to a secure, stable release. Post-upgrade, it launches regression test suites to detect breakages, applies required code patches, and flags any unresolved compatibility issues. A human-readable changelog is auto-generated, listing updated packages, reasons for change (e.g., vulnerability fix, feature parity), and impacted modules. This workflow ensures software supply chain hygiene and aligns with security best practices such as SBOM (Software Bill of Materials) generation.

\item \textbf{Regression Bug Fixing}

In enterprise-grade pipelines, where minimizing downtime is critical, agentic systems provide rapid response mechanisms for resolving regressions. Instructed with “Identify and fix any regression bugs introduced in the last release,” the agent fetches the latest commits, runs test pipelines, and maps failures to code changes using blame heuristics or statistical fault localization techniques. Upon identifying root causes, it proposes targeted patches and verifies fixes through retesting. If successful, the fix is committed with rollback metadata. This not only reduces mean time to resolution (MTTR) but also minimizes human debugging cycles during post-deployment phases.

\item \textbf{CI/CD Pipeline Automation}

Setting up and maintaining CI/CD pipelines is essential but repetitive ideal for delegation to agentic systems. When asked to “Set up and maintain a CI/CD pipeline that builds, tests, and deploys our microservices to AWS,” the agent scaffolds GitHub Actions or GitLab CI YAML files, configures secrets management (e.g., via AWS IAM), builds Docker containers, and deploys them to ECS or Lambda environments. It also implements rollback triggers, environment matrix testing, and artifact versioning. This workflow transforms DevOps setup into a task-oriented scriptable process, freeing engineers to focus on application-level issues.

\item \textbf{Automated Security Auditing}

Agentic coding tools are highly effective for security auditing. Given the prompt “Scan the codebase for OWASP Top 10 vulnerabilities, apply fixes, and generate a security report,” the agent runs static analysis (e.g., CodeQL, Bandit), applies sanitization patches (e.g., escaping input fields), and produces a PDF security report complete with fix diffs, severity scores, and coverage metrics. It also enforces policy-as-code integrations to flag future regressions. This is crucial for SOC 2 or GDPR-compliant organizations requiring continuous assurance mechanisms.

\item \textbf{Large-Scale Code Migration}

Legacy code migrations, such as “Migrate the codebase from Python 2.7 to Python 3.x,” are complex and error-prone. The agent tokenizes source files into abstract syntax trees (ASTs), applies rule-based transformations (e.g., print statements to functions, Unicode updates), and updates dependency management (e.g., replacing \texttt{pip2} libraries). After transformation, it runs and validates unit and integration tests, logging all conversions for developer audit. This use case illustrates how agentic systems can bridge large syntactic gaps and reduce organizational technical debt with traceable automation.

\item \textbf{Automated Documentation Generation}

Documentation often lags development. Agentic systems solve this via prompts like “Generate API documentation for all endpoints, including usage examples and parameter descriptions.” The agent extracts function docstrings, converts them into OpenAPI-compliant specs, and deploys an interactive Swagger UI. If inline docs are missing, it infers descriptions from usage patterns or model inference. The documentation is versioned with the codebase and updated on every relevant commit. This ensures code maintainability, simplifies onboarding, and aligns with industry standards such as REST maturity models or GraphQL schema introspection.

\item \textbf{Performance Optimization}

Performance profiling and optimization tasks are ideal candidates for agentic workflows. Given “Profile the application, identify bottlenecks, and optimize slow database queries,” the agent instruments performance probes using \texttt{cProfile}, \texttt{perf}, or Chrome DevTools depending on stack. It locates hotspots such as nested loops or unindexed queries, applies refactors (e.g., SQL indexing, memoization, pagination), and verifies performance gains with benchmarks. The final report includes before/after metrics, call graphs, and optimization rationale. This use case demonstrates how agentic coding tools can enforce non-functional requirements through measurable performance metrics.

\item \textbf{End-to-End Feature Implementation}

Agentic coding systems are capable of implementing complex, multi-component features. For instance, the prompt “Implement a new payment gateway integration, update the UI, backend, and database, and ensure all workflows are tested” triggers code generation for frontend forms (e.g., Stripe.js), backend API routes, database schema updates, and test coverage via Cypress or Postman. The agent updates configuration files, manages secrets, and deploys to staging environments. Human developers review logs, inspect schema diffs, and approve PRs. This end-to-end autonomy illustrates the promise of agentic systems in full-stack development.

\item \textbf{Automated Rollback and Recovery}

For production environments, agentic systems serve as first responders. Prompted with “Monitor production for critical errors, and if detected, automatically roll back to the last stable version and notify the team,” the agent uses observability tools (e.g., Datadog, Sentry) to watch logs for error spikes. If conditions match a critical failure signature, the agent initiates rollback via GitOps workflows (e.g., ArgoCD), verifies system health post-rollback, and dispatches alerts via Slack or email. This application reduces incident response time and adds resilience to mission-critical deployments.

\end{enumerate}

\section{Industry Trends and Convergence}

The evolving interface between humans and artificial intelligence in software development has given rise to two prominent paradigms: Vibe Coding and Agentic Coding. Originally designed with distinct purposes  vibe coding as an exploratory, conversational mode and agentic coding as a structured, autonomous execution model  these approaches are increasingly converging. This convergence is not coincidental; it reflects broader socio-technical demands across domains such as enterprise automation, developer education, and creative software innovation. In this section, we examine the emerging hybrid architectures, adoption trajectories across industry sectors, and the synthesis of best practices that signal a maturing AI-assisted development ecosystem.

\subsection{Emergence of Hybrid Models}

Contemporary platforms are beginning to blur the once-clear lines between reactive conversational assistants and autonomous agent frameworks. Vibe coding systems, initially confined to prompt-based generation in natural language interfaces, have started incorporating execution capabilities, persistent context, and basic planning modules. For instance, tools like Replit Ghostwriter now support inline execution and debugging, offering partial autonomy within conversational workflows.

Conversely, agentic platforms such as OpenAI Codex, Claude Code, and Google Jules have introduced interface elements from vibe coding: accepting high-level natural language goals, providing step-by-step feedback, and engaging in clarifying dialogue. These developments illustrate an architectural fusion in which conversational flexibility is paired with autonomous execution, leading to hybrid systems capable of decomposing, planning, validating, and summarizing multi-step software tasks.

Hybrid models allow users to issue abstract objectives (e.g., “Build a secure login system with 2FA and audit logging”), which are parsed by the AI into discrete submodules. The agent then executes each step, verifies the results via tests, and presents logs and artifacts for review. This synthesis offers three distinct benefits: (i) conversational speed in ideation, (ii) execution precision with agent control, and (iii) a continuous loop of refinement via real-time feedback. However, challenges remain in ensuring explainability, safe prompt handling, and seamless cross-platform integration.

\subsection{Enterprise and Educational Adoption}

The convergence of AI coding paradigms is not only theoretical but increasingly visible in industry and education. Agentic systems are gaining traction in enterprise environments due to their capacity for automating mission-critical tasks. Organizations such as Cisco employ agentic frameworks for regression testing, legacy code refactoring, and continuous integration workflows. Similarly, Kodiak Robotics leverages agentic tools for safety-critical verification in autonomous driving software.

In contrast, vibe coding is widely adopted in education and individual development. Platforms such as VS Code and Replit embed vibe-oriented coding assistants directly into the IDE, allowing students and solo developers to explore new APIs, build prototypes, and debug through conversational interaction. Coding bootcamps use these systems for instructional scaffolding  providing code suggestions, explanations, and project feedback.

Adoption patterns exhibit a dual structure: top-down implementation of agentic systems in enterprise pipelines, and bottom-up adoption of vibe tools among independent users. Despite their promise, adoption faces three main barriers: (i) governance concerns around AI decision transparency and security, (ii) skepticism toward black-box automation among seasoned developers, and (iii) the need for retraining teams in AI-centric workflows and agent supervision.

\subsection{Balanced Development Practices}
As these paradigms continue to merge, a balanced model of human-AI collaboration is emerging. In this paradigm, developers use vibe interfaces to articulate system intent (e.g., “Design a multilingual registration form with spam filters”), and agents execute subcomponents  backend validation, frontend form generation, and anti-spam logic  under the developer’s supervision. The human then reviews and refines outputs, enforces policy compliance (e.g., GDPR), and initiates deployment.

Balanced practices offer the best of both paradigms: the creative freedom and speed of vibe coding, and the repeatability, quality assurance, and architectural rigor of agentic systems. Empowered by this synergy, non-programmers can initiate software logic via natural language, while engineers maintain control over architecture, policy, and integration. Still, three unresolved challenges remain: (i) ensuring runtime security against emerging prompt-based or model-exploitation vulnerabilities, (ii) implementing comprehensive and interpretable audit trails for AI decisions, and (iii) preserving and cultivating developer expertise in the face of rising abstraction and automation.

The convergence of vibe and agentic coding represents a paradigmatic shift in AI-assisted software engineering. Forward-looking organizations are embracing hybrid workflows that leverage intuitive ideation and autonomous execution. Those that invest in explainability, modular agent design, and developer empowerment are likely to lead the next era of resilient and scalable software innovation.

\section{Challenges and Limitations}

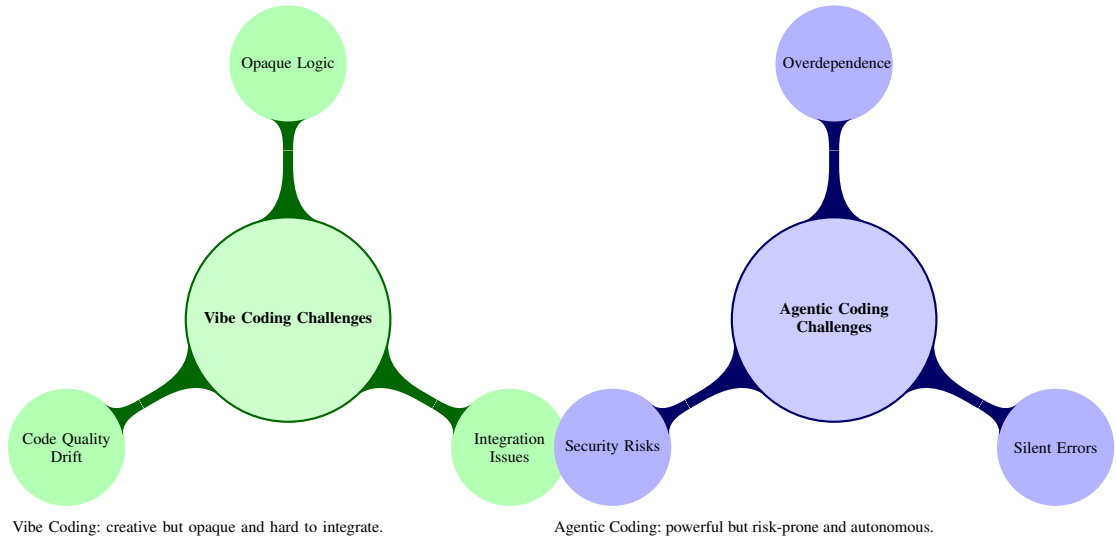
\begin{figure*}[htbp]
\centering
\scalebox{0.68}{
\begin{minipage}[t]{0.45\textwidth}
\centering
\begin{tikzpicture}
  \path[mindmap, text=black, concept color=green!40!black,
        level 1 concept/.append style={sibling angle=120}]
    node[concept, font=\bfseries\small, fill=green!20] {Vibe Coding Challenges}
      [clockwise from=90]
      child { node[concept, concept color=green!30!white] {Opaque Logic} }
      child { node[concept, concept color=green!30!white] {Integration Issues} }
      child { node[concept, concept color=green!30!white] {Code Quality Drift} };
\end{tikzpicture}
\captionof*{figure}{\small Vibe Coding: creative but opaque and hard to integrate.}
\end{minipage}
\hfill
\hspace{3 cm}
\begin{minipage}[t]{0.45\textwidth}
\centering
\begin{tikzpicture}
  \path[mindmap, text=black, concept color=blue!40!black,
        level 1 concept/.append style={sibling angle=120}]
    node[concept, font=\bfseries\small, fill=blue!20] {Agentic Coding Challenges}
      [clockwise from=90]
      child { node[concept, concept color=blue!30!white] {Overdependence} }
      child { node[concept, concept color=blue!30!white] {Silent Errors} }
      child { node[concept, concept color=blue!30!white] {Security Risks} };
\end{tikzpicture}
\captionof*{figure}{\small Agentic Coding: powerful but risk-prone and autonomous.}
\end{minipage}
}
\caption{Comparison of challenge domains for Vibe Coding (left) and Agentic Coding (right) using mindmap representation.}
\label{fig:vibe_agentic_mindmap}
\end{figure*}

Despite their transformative potential, both Vibe Coding and Agentic Coding present critical limitations as illustrated in Figure \ref{fig:vibe_agentic_mindmap} that must be understood for safe deployment, sustainable adoption, and long-term developer resilience. These challenges are architectural, procedural, and cognitive in nature arising not only from technical immaturity but also from systemic gaps in explainability, oversight, and security. This section provides a detailed analysis of the emerging risks and constraints associated with both paradigms, grounded in real-world case studies and interdisciplinary insights from software engineering, human-computer interaction, and AI ethics.

\subsection{Limitations of Agentic Coding}

Agentic coding systems, while promising high degrees of autonomy, introduce risks that arise from reduced human oversight, opaque execution logic, and uncontrolled access to critical infrastructure. One of the most pressing concerns is the overdependence on agents for routine and high-stakes engineering tasks. As developers become increasingly reliant on autonomous systems, their engagement with core programming concepts and debugging strategies may diminish, leading to skill atrophy and reduced situational awareness. This is analogous to findings in aviation automation and clinical decision-support systems, where passive user roles have been shown to degrade cognitive vigilance. The long-term consequence in software engineering could be a workforce that is poorly equipped to intervene during edge-case failures or system crises.

Another serious concern is the potential for silent error propagation. Agentic systems operating across multiple modules can introduce logic faults or regressions that go undetected until deployment. Because these agents modify code, adjust configurations, and interface with APIs at runtime, a fault introduced in one subsystem can cascade downstream particularly if agents are not equipped with rollback mechanisms or observability hooks. Examples include global refactors that destabilize microservice communication protocols or schema changes that disrupt dependent services. Robust mitigation requires explainable agent decisions, real-time anomaly detection, and strict version control governance.

In addition, the expanded runtime privileges of agentic platforms create new vectors for security vulnerabilities. Autonomously acting agents may unwittingly expose sensitive data, mishandle authentication tokens, or install unverified dependencies. Threats such as prompt injection, dependency confusion, or secret leakage via AI-generated commits are increasingly documented in agentic pipelines. Defending against these vulnerabilities necessitates rigorous sandboxing, zero-trust security policies, prompt sanitization, and cryptographic verification for all actions taken by autonomous code agents.

\subsection{Limitations of Vibe Coding}

While vibe coding tools promote flexibility and creative exploration, they suffer from systemic challenges rooted in the opacity of model outputs and the lack of integration with formal software development lifecycles. Chief among these is the black-box nature of generation. Most LLM-based coding assistants do not expose their internal decision processes, making it difficult for developers to validate code correctness, interpret logic decisions, or trace performance regressions. This undermines trust in high-stakes domains, especially when generated code is inserted into production pathways. Furthermore, the stochasticity of model outputs can lead to inconsistent quality even under near-identical prompts.

Another prominent limitation of vibe coding is its poor compatibility with production-oriented development systems. Generated code often functions well in isolation but fails when incorporated into real-world environments due to missing context such as authentication flows, deployment configurations, or CI/CD hooks. Without access to full project state or execution context, LLMs are prone to suggesting solutions that ignore runtime dependencies or system architecture constraints. This makes them ideal for scaffolding or ideation, but suboptimal for system-level implementation unless paired with structured review protocols and toolchain integration.

Finally, the rapid, iterative style of vibe coding can erode long-term code quality. Developers focused on short feedback cycles may forgo documentation, unit testing, or adherence to architectural principles. Over time, this contributes to codebases riddled with duplication, inconsistent naming, security shortcuts, and unmaintainable logic an accumulation of technical debt with systemic consequences. Effective interventions include mandatory linting, automated test scaffolding, and enforced review pipelines for all AI-assisted code merges. Vibe tools should serve as accelerators, not replacements, for engineering best practices.

\section{Future Roadmap: Advancing Agentic AI for Autonomous Software Engineering}
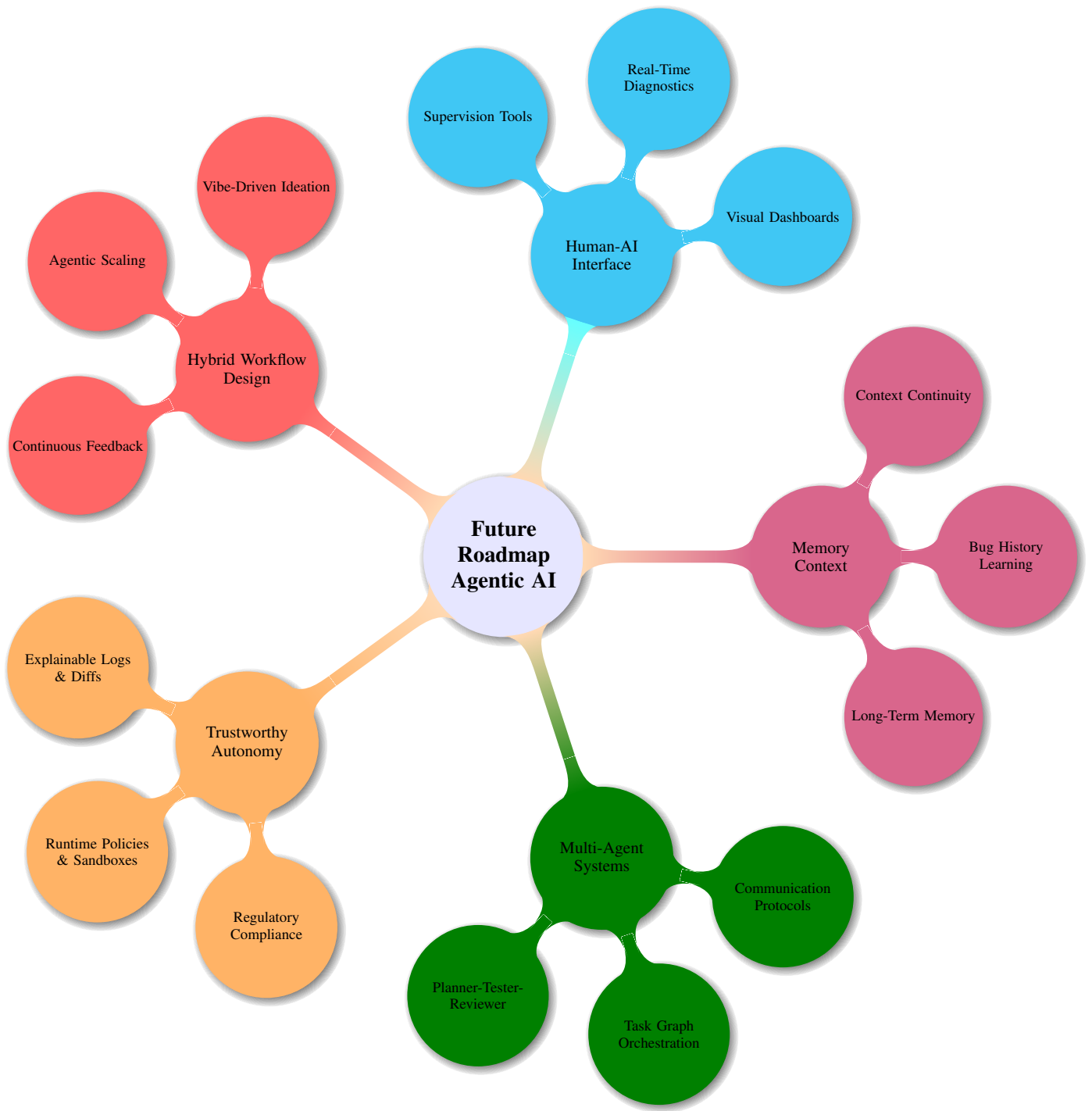
\begin{figure*}[htbp]
\centering
\scalebox{0.90}{
\begin{tikzpicture}[
  mindmap,
  grow cyclic,
  concept color=orange!30,
  every node/.style={concept, circular drop shadow, minimum size=2.5cm, text width=2.5cm, align=center},
  root concept/.style={concept color=blue!10, font=\large\bfseries, minimum size=3cm},
  level 1/.append style={level distance=6cm, sibling angle=72},
  level 2/.append style={level distance=3.5cm, sibling angle=60}
]

\node[root concept] {Future Roadmap\\Agentic AI}
  child[concept color=orange!60] { node {Trustworthy Autonomy}
    child { node {Explainable Logs\\\& Diffs} }
    child { node {Runtime Policies\\\& Sandboxes} }
    child { node {Regulatory Compliance} }
  }
  child[concept color=green!50!black] { node {Multi-Agent Systems}
    child { node {Planner-Tester-Reviewer} }
    child { node {Task Graph Orchestration} }
    child { node {Communication Protocols} }
  }
  child[concept color=purple!60] { node {Memory \\ Context}
    child { node {Long-Term Memory} }
    child { node {Bug History Learning} }
    child { node {Context Continuity} }
  }
  child[concept color=cyan!60] { node {Human-AI Interface}
    child { node {Visual Dashboards} }
    child { node {Real-Time Diagnostics} }
    child { node {Supervision Tools} }
  }
  child[concept color=red!60] { node {Hybrid Workflow Design}
    child { node {Vibe-Driven Ideation} }
    child { node {Agentic Scaling} }
    child { node {Continuous Feedback} }
  };

\end{tikzpicture}
}
\caption{Mindmap overview of the future roadmap for Agentic AI in autonomous software engineering, including key directions such as trustworthy autonomy, multi-agent systems, hybrid workflow integration, memory persistence, and human-AI supervision.}
\label{fig:agentic_ai_mindmap}
\end{figure*}

The future of AI-assisted programming (As depicted in Figure \ref{fig:agentic_ai_mindmap}) will be increasingly shaped by the maturity and proliferation of agentic coding systems platforms that do not merely assist in code generation but autonomously plan, execute, test, and validate software development tasks across the engineering lifecycle. As organizations seek to scale automation, reduce technical debt, and manage complex digital ecosystems, agentic AI stands at the frontier of practical transformation. This roadmap outlines the core trajectories, challenges, and infrastructure required to operationalize agentic systems responsibly and at scale.

\subsection{Architecting Trustworthy Autonomy}

The next generation of agentic AI must prioritize trust, reliability, and governance. This entails a shift from static model inference to dynamic, feedback-rich execution environments. Agents must be designed with embedded explainability generating transparent logs, semantic diffs, decision traces, and rollback records. As software teams integrate agents into CI/CD pipelines, static and dynamic analysis tools must be extended to interpret AI-generated logic and expose risks early.

Moreover, agentic systems must comply with software assurance standards. This includes regulatory compliance (e.g., GDPR, ISO/IEC 27001), organizational policies (e.g., coding conventions, security models), and runtime safety guarantees. Future agentic frameworks will require built-in guardrails such as rule-based policy engines, automated rollback triggers, and runtime permission sandboxes that enforce zero-trust principles during execution.

\subsection{Multi-Agent Collaboration and Specialization}

Scalability in agentic coding will emerge not from a single monolithic agent, but from a constellation of specialized sub-agents planners, coders, testers, reviewers coordinated by an orchestrator. Inspired by distributed systems theory and modular programming paradigms, such multi-agent architectures will enable parallel task decomposition, resource optimization, and resilience through redundancy.

To enable meaningful collaboration among agents, a shared language and structured communication protocol will be necessary. Advancements in function-calling, task graph serialization, and contextual memory sharing will allow agents to synchronize states, pass artifacts, and coalesce outputs into consistent deliverables. This architectural pattern will mirror human software teams, enabling software construction to scale without linear increases in human supervision.

\subsection{Memory, Context, and Long-Term Adaptation}

Agentic AI will only succeed in production settings if it can reason across time, projects, and usage contexts. Future systems must integrate both short-term (working) memory and persistent memory (organizational preferences, historical codebase patterns, bug history). Memory-augmented LLMs or retrieval-based hybrid agents will be critical in maintaining task continuity and avoiding context fragmentation over multi-hour or multi-day tasks.

Additionally, learning from operational feedback will become central to agent refinement. Mechanisms such as reinforcement learning from human feedback (RLHF), offline evaluation from logs, and interactive model distillation will allow agents to align with evolving team practices, technology stacks, and user expectations. These capabilities will gradually shift agents from static models to continuously improving team members.

\subsection{Human-AI Collaboration Infrastructure}

Agentic coding should not replace developers but elevate them to higher-order roles strategic planners, architectural reviewers, and AI supervisors. To support this shift, integrated human-agent interfaces must evolve. Rich visualization dashboards, interpretability overlays, interactive agent simulations, and real-time progress diagnostics will empower humans to supervise AI workflows effectively.

Training developers to interpret, configure, and intervene in agent behavior will be essential. AI literacy programs, sandbox testing environments, and debugging toolkits tailored for AI-generated systems will form the backbone of future software education and organizational readiness.

\subsection{5. Strategic Integration and Hybrid Workflow Design}

The future of software development lies not in choosing between vibe coding and agentic coding but in combining their strengths. Vibe coding ideal for early-stage ideation, UX design, and experimental workflows will serve as the creative front-end. Agentic coding engineered for precision, automation, and long-horizon planning will operationalize and scale those ideas into robust, production-grade systems.

Hybrid workflows will increasingly rely on seamless transitions: vibe tools initiating conceptual drafts, agentic agents refining and deploying them, and human teams orchestrating this interplay through continuous feedback loops. These workflows will not only maximize efficiency and innovation but also create resilient software systems that adapt to future complexity.

Agentic AI promises a paradigm shift in software engineering transforming AI from a passive assistant to an autonomous co-developer. Realizing this potential demands more than algorithmic power; it requires trustworthy infrastructure, human-centered design, and rigorous governance. The roadmap to agentic maturity is a socio-technical journey one that redefines collaboration, responsibility, and intelligence in software creation. Those who invest early in this convergence will shape the foundational tools of the next engineering era.

\subsection{Historical Evolution of AI Agents: From Rule-Based Systems to Agentic AI}

The trajectory of AI agents reflects a four-decade-long transformation from symbolic, rule-based automation to generative, goal-directed intelligence. Understanding this historical evolution is essential for grounding future developments in Agentic AI within the broader arc of artificial intelligence research.

In the 1990s, AI agents were largely constructed as \textit{symbolic software entities} \cite{eriksson1999symbolic, bradshaw1997introduction, fenton2002software}, grounded in deterministic logic \cite{abiteboul1991non, graf1986logic} and finite-state control systems \cite{wagner2006modeling, drumea2004finite}. Representative systems included intelligent tutoring agents like \textit{SHERLOCK} \cite{katz1992modelling, preece2017sherlock, katz2020sherlock} and \textit{Andes} \cite{vanlehn2005andes, schulze2000andes, gertner2000andes}, which delivered domain-specific instruction using scripted pedagogical rules. Concurrently, mobile agents such as General Magic’s \textit{Telescript} \cite{chess1996mobile} facilitated lightweight task execution across distributed networks \cite{milojicic1999mobile}. Behavioral models like the Belief-Desire-Intention (BDI) framework (e.g., PRS, JAM) formalized rational planning under constrained environments \cite{kardas2018domain}. However, these early agents lacked learning capability, contextual reasoning, and autonomy beyond their initial programming.

The 2000s and 2010s marked a pivotal shift toward \textit{learning-based and networked agents} \cite{gelenbe2001simulation, arel2010reinforcement}. Advances in reinforcement learning (e.g., TD-Gammon\cite{tesauro2002programming} and Deep Q-Networks) enabled agents to optimize behavior through reward-driven feedback. Multi-agent systems (MAS) using frameworks such as JADE enabled collaboration  negotiation, and task allocation across agents\cite{vitabile2009extended, bellifemine2001developing, bellifemine2008jade}. These architectures found use in traffic systems, robotic swarms, and industrial simulation. Chatbots and NLP agents matured with systems like Siri and Alexa, offering human-AI dialogue capabilities grounded in statistical language models. While more adaptable, these systems remained bounded by task-specific learning and lacked the emergent planning and reasoning evident in human cognition.

The 2020s usher in a new paradigm: \textit{agentic coding powered by large language models (LLMs)}. Unlike their predecessors, LLM-based agents (e.g., AutoGPT \cite{yang2023auto}, BabyAGI \href{https://github.com/yoheinakajima/babyagi}{Baby AGI}, Devin \href{https://devin.ai/}{Devin AI}, Codex \href{https://openai.com/index/openai-codex/}{OpenAI Codex}) demonstrate the ability to decompose abstract goals, synthesize code, invoke APIs, interact with development environments, and reason iteratively through planning-execution-feedback loops. These agents use components such as memory buffers, tool use modules, sandboxed execution environments, and self-verification routines to operate autonomously in complex, real-world software engineering tasks. For instance, Codex-based systems execute end-to-end pipelines  cloning GitHub repositories, updating codebases, testing, and committing patches  without line-by-line developer supervision. This level of autonomy and contextual awareness signals a profound leap in agent capability.

These trends signify not only an evolution in computational architecture but a deepening of the cognitive model underlying artificial agents. The agent is no longer a function executor or rule follower, but a self-reflective, goal-seeking entity capable of autonomous software generation, debugging, and deployment. As we move forward, the synthesis of historical rule-based robustness with modern generative flexibility offers a foundation for building trustworthy, transparent, and socially-aligned Agentic AI systems. By drawing on decades of agent modeling  from BDI frameworks to LLM orchestration  future systems can achieve both operational excellence and epistemic alignment with human objectives.

\section{Conclusion}
This review presented a comprehensive comparative analysis of two emerging paradigms in AI-assisted software development: vibe coding and agentic coding. By systematically tracing their conceptual foundations, interaction models, technical architectures, and practical workflows, we highlighted how these approaches reflect distinct philosophies in human-AI collaboration. Vibe coding, as discussed, is rooted in intuition-driven, prompt-based development that emphasizes co-creation, creativity, and flexibility. It foregrounds developer control, iterative experimentation, and conversational ideation, supported by a set of foundational cognitive skills that reorient programming as a semiotic dialogue between human intent and machine synthesis. 

In contrast, agentic coding represents a paradigm shift toward autonomous software engineering systems. These systems leverage architectural modules such as planners, executors, memory buffers, and feedback-integrated agents capable of operating across the software lifecycle. Our investigation into their execution models, safety protocols, and developer interaction revealed how agentic systems enable delegation of complex development tasks while introducing new challenges in explainability, governance, and secure operation. We further compared their autonomy mechanisms, debugging strategies, tool ecosystems, and implementation pathways. Through illustrative workflows and real-world applications  from social media prototyping to automated legacy refactoring  we demonstrated how vibe coding excels in ideation and learning environments, while agentic coding is well-suited for scalable automation, CI/CD integration, and enterprise-grade reliability. The final sections synthesized industry trends, adoption barriers, and the convergence toward hybrid workflows, where conversational ideation is seamlessly fused with autonomous implementation. 

We also outlined a future roadmap for advancing agentic AI  emphasizing trustworthy autonomy, multi-agent orchestration, contextual memory, and human-centered interfaces. Ultimately, this review frames vibe and agentic coding not as competing paradigms, but as complementary pillars in the future of AI-driven software engineering  each contributing uniquely to the evolving landscape of human-machine co-development.

\section*{Acknowledgement} This work was supported by the National Science Foundation and the United States Department of Agriculture, National Institute of Food and Agriculture through the ``Artificial Intelligence (AI) Institute for Agriculture” Program under Award AWD003473, and AWD004595, Accession Number 1029004, "Robotic Blossom Thinning with Soft Manipulators". The publication of the article in OA mode was financially supported by HEAL-Link.
\normalsize
\section*{Declarations}
The authors declare no conflicts of interest.

\section*{Statement on AI Writing Assistance}
ChatGPT and Perplexity were utilized to enhance grammatical accuracy and refine sentence structure; all AI-generated revisions were thoroughly reviewed and edited for relevance. Additionally, ChatGPT-4o was employed to generate realistic visualizations.

\bibliography{references.bib}

\bibliographystyle{ieeetr}

\end{document}